%% file: hd-eavesdropping.tex
\documentclass[11pt]{article}
\usepackage[margin=1in]{geometry}

\usepackage{import}
\usepackage[written-dimensions,show-todos,highlight-cmd,highlight-ref,final-draft]{hd-eavesdropping}

\title{High-Dimensional Quantum Eavesdropping: A Hypothetical Attack on BB84 \& SSP}
\author{Christopher Dunne}
\date{December 2, 2023}

\begin{document}
    \nocite{Heisenberg1983}  % Not citing directly because its translated. Use \citeHUP instead

    \listoftodos[Notes]
    \pagebreak
    \newlogic{Create a high-dimensional variant of BB84 (HD-BB84) abbreviation.}
    \newlogic{Create a quantum bit (qubit) abbreviation.}
    \needtofix[inline]{Fix funky spacing in references and paper. Make everything left aligned.}
    \needtofix[inline]{Tables aren't centerd in results section.}
    \expandupon[inline]{7/23/24: Explain quantum sensors/elaborate on them. Attack is same but w different sensor.}
    \topic{Abstract/Conclusion: Should wait to implement QKD algorithms?}
    \repetitive[inline]{Change `novel attack strategy`.}

    % Paper
    \maketitle
    \pagebreak

    % Abstract
    \import{sections/abstract/}{abstract.tex}

    \pagebreak

    % Table of Contents
    \tableofcontents
    \pagebreak

    % Sections
    \import{sections/introduction/}{introduction.tex}

    \pagebreak

\import{sections/lit-review/}{lit-review.tex}

%    \clearpage

\import{sections/methodology/}{methodology.tex}

    \clearpage

\import{sections/results/}{results.tex}
    \clearpage

\import{sections/discussion/}{discussion.tex}

    \clearpage

    \clearpage
    \import{sections/conclusion/}{conclusion.tex}

    % References
    \clearpage
    \bibliographystyle{apacite}
    \bibliography{hd-eavesdropping}

    % Appendices
    \clearpage
    \appendix
    \section{Reading Qudit Measurements in HD-Eavesdropping} \label{app:res-format}
    \import{appendices/}{result-format.tex}
    \clearpage
    \section{Conversion Matrices} \label{app:conv-unitaries}
    \import{appendices/}{conv-unitaries.tex}

\end{document}

%% file: sections/abstract/abstract.tex
% 100 - 300 words
% https://www.scribbr.com/dissertation/abstract/
% Current Word Count: 106 (7/23/24)
% https://writing.wisc.edu/handbook/assignments/writing-an-abstract-for-your-research-paper/
\begin{abstract}
    Quantum key distribution algorithms are considered secure because they leverage quantum phenomena to provide
    security.
    As such, eavesdroppers can be detected by analyzing the error rate in the shared key obtained by the parties
    performing the key exchange.
    Nevertheless, this paper developed and investigated a novel attack strategy capable of being undetected through
    error analysis of the shared key.
    Said attack entails an eavesdropper measuring the quantum channel (i.e., the channel used to transmit
    quantum particles between two parties) using a higher dimension than that used by the legitimate parties.
    By measuring a particle in a dimension that spans all possible states in a given quantum key distribution algorithm,
    the collapsed state of the measured particle should be undetectable when measured in the lower dimension.
    % Step 1 - Introduction
%    This paper investigates a novel attack strategy on quantum key distribution algorithms wherein an eavesdropper
%    measures the quantum channel using a higher dimension than that used by the legitimate parties.
    % Step 2 - Methods
    To analyze the proposed high-dimensional eavesdropping attack, the simulator Cirq was used to model the attack on
    BB84, a \ndimensional{4} variant of BB84, and \glsentrylong{ssp}.
    % Step 3 - Results
    The results of these simulations show the efficiency of this attack, with an eavesdropper being undetectable
    through \glsentrylong{qber} analysis in each algorithm
    % Step 4 - Discussion
    This promotes the need for further analysis of the high-dimensional eavesdropping attack on physical hardware and
    other quantum key distribution algorithms.
    \expandupon[inline]{Doing so could prevent...?}
\end{abstract}

%% file: sections/introduction/introduction.tex
%! Author = chris
%! Date = 6/28/2024
% See ARB

\section{Introduction} \label{sec:introduction}

The \gls{qkd} algorithm BB84 leverages Heisenberg's uncertainty principle and the no-cloning theorem to provide
security and serves as the basis for several other \gls{qkd} algorithms~\cite{Sabani2022}.
By encoding data in two non-orthogonal quantum states, one creates a secure quantum channel that cannot reliably be
read or copied by an eavesdropper without altering it in such a way that will be easily noticed~\cite{Bennett1984}.
However, this assumes that an eavesdropper is measuring the quantum channel using the same number of dimensions as that
used by the legitimate parties.
This paper will explore the usage of higher dimensions to eavesdrop BB84 and several variants based on it.
This high-dimensional eavesdropping attack was modeled using the quantum simulator Cirq.

\subimport{subsections}{background.tex}

\subimport{subsections}{problem-statement.tex}

\subimport{subsections}{significance.tex}
\subimport{subsections}{nature-of-study.tex}

\subimport{subsections}{framework.tex}

\subimport{subsections}{hypotheses.tex}
% Definitions
\subimport{subsections}{assumptions.tex}

%% file: sections/introduction/subsections/background.tex
\subsection{Background of Study} \label{subsec:introduction-background}

Cryptography is the practice of protecting data by hiding its content, preventing unauthorized access, and preventing
undetected modification~\cite{Franklin2020, Amellal2023}.
It is utilized by billions of devices to secure sensitive information such as financial and medical
records~\cite{Shingari2024} and is vital in the operation of cloud computing~\cite{Sasikumar2024} and the
internet~\cite{Liu2024}.
There are two types of encryption algorithms, symmetric and asymmetric algorithms.
Symmetric algorithms use the same key for both encryption and decryption, whereas asymmetric algorithms use
different keys for these processes~\cite{Sood2023}.
Secure key distribution is essential in both types of algorithms.

Key distribution entails the transportation of cryptographic keys and other keying material between a party that has a
key and a party that needs a key.
This process is critical to ensure the security of both symmetric and asymmetric encryption
algorithms~\cite{Barker2020}.
Classical key distribution algorithms rely on mathematically difficult to solve problems to provide security, whereas
\gls{qkd} algorithms rely on the fundamental laws of nature~\cite{Singh2022, Lee2022, Bommi2023}.

%% file: sections/introduction/subsections/problem-statement.tex
\subsection{Problem Statement} \label{subsec:introduction-problem-statement}
% https://www.enago.com/academy/research-problem-statement/#:~:text=A%20problem%20statement%20describes%20the,research%20project%20aims%20to%20fill.

It is easy to assume that \gls{qkd} algorithms provide unconditional security due to their leverage of quantum
mechanics.
Despite this, it is crucial to thoroughly explore all possible attack vectors on an algorithm before implementation.
This paper develops and investigates a novel attack strategy that exploits quantum phenomena to eavesdrop on the
\gls{qkd} algorithm BB84 and two of its derivatives.

The Heisenberg uncertainty principle states that it is impossible to measure a particle without altering its state.
This is utilized by \gls{qkd} algorithms to provide security as an eavesdropper will always produce a change in the
system~\cite{Singh2022}.
However, this is reliant on the assumption that the eavesdropper measures the quantum particle in the same number
of dimensions as that used by the legitimate parties.
This assumption will be challenged by a proposed high-dimensional eavesdropping attack to demonstrate a potential
weakness in \gls{qkd} algorithms that rely on the uncertainty principle to provide security.

%It is easy to assume that \gls{qkd} algorithms provide unconditional security because they leverage quantum mechanics.
%Despite this, it is important to thoroughly explore all possible attack vectors on an algorithm before implementation.
%This paper will explore a novel attack strategy that leverages quantum phenomenon to attempt an attack on the \gls{qkd}
%algorithm BB84 and two other algorithms based on BB84.
%
%The Heisenberg uncertainty principle states that it is impossible to measure without changing the state of said
%particle.
%This is leveraged by \gls{qkd} algorithms to provide security as an eavesdropper will always produce a change in the
%system~\cite{Singh2022}.
%However, this is reliant on the assumption that the eavesdropper measures the quantum particle in the same number
%of dimensions as that used by the legitimate parties.
%A proposed attack strategy on \gls{qkd} algorithms that challenges this assumption will be analyzed to demonstrate a
%potential weakness in \gls{qkd} algorithms that rely on the uncertainty principle to provide security.

%% file: sections/introduction/subsections/significance.tex
\subsection{Significance of Study} \label{subsec:introduction-significance}
%\expandupon[inline]{7/23/24: Add more examples and elaborate on SIKE attack vector.}

% SIKE
Any algorithm intended to protect devices must undergo rigorous analysis before implementation.
The necessity of doing so can be seen in the \gls{sike} algorithm, which was a final candidate that \gls{nist} was
evaluating for its post-quantum cryptography standards.
However, \citeauthor{Castryck2023} demonstrated an attack based on a theorem first proposed in 1997 that was capable of
breaking \gls{sike} in approximately 62 minutes using a legacy system~\cite{Castryck2023}.
Before this attack had been demonstrated, \gls{sike} was widely considered to be a secure cryptographic
algorithm~\cite{Longa2021, Costello2021, Stratil2021}.

% Rainbow
Another algorithm that \gls{nist} was evaluating for its post-quantum security was the Rainbow signature scheme.
Rainbow was first proposed in 2005 by \citeauthor{Ding2005} and is a multivariate signature scheme based on the unbalanced
Oil and Vinegar signature scheme~\cite{Ding2005}.
In 2022, \citeauthor{Beullens2022} found an attack vector capable of breaking Rainbow in a week on a laptop.
To secure the Rainbow algorithm against this attack, one would have to increase the size of the keys and signatures
generated.
However, doing so would make it less efficient than the Oil and Vinegar algorithm it was built upon.
Furthermore, further investigation would still be required to see if the attack could be optimized for the larger
key sizes of this modified version of Rainbow ~\cite{Beullens2022}.

% Importance of Crypto
Cryptographic algorithms are utilized by a wide spectrum of areas such as healthcare, banking, retail, and government
entities~\cite{Singh2022, Kose2024}.
A premature implementation of \gls{sike} or Rainbow before either of these attacks were discovered would have severe
consequences for any of said entities.
As such, analyzing attack vectors on cryptographic algorithms before mass adoption is important to prevent the loss
of vital information and resources.

Finally, multiple regulations require cryptography.
Two such regulations are \gls{gdpr} and \gls{hipaa}.
The \gls{gdpr} is applicable to all organizations that collect personal data in the European Union~\cite{CEU2016},
whereas \gls{hipaa} is applicable solely to sensitive health information of patients in the United
States~\cite{OFR1996}.

%TODO: Use AI attack on Kyber to demonstrate import of attacks that don't break alg to such a degree.
% Find a source discussing the importance of finding and analyzing attacks on cryptographic algorithms.

% More recent example - AI attack on Kyber (though not nearly as damning)

% QKD algorithms must be thoroughly tested before implementation <CITE>
% Provide explanations/examples of why it is important.
% Example - PQC Algorithm that was immediately broken?
% Will be aprroaching a new attack strategy

% Importance of Key Exchange:
% https://ieeexplore.ieee.org/abstract/document/10396317
% Discuss what wouldve happened if vulnerability was not discovered before mass adoption/replacement of current standards

% https://www.quantamagazine.org/post-quantum-cryptography-scheme-is-cracked-on-a-laptop-20220824/
% Hence why competitions like NIST’s are so important. In the previous round of the NIST competition, Ward Beullens, a cryptographer at IBM, devised an attack that broke a scheme called Rainbow in a weekend
% Look into Rainbow break

%% file: sections/introduction/subsections/nature-of-study.tex
% See research method of ARB

\subsection{Nature of Study} \label{subsec:introduction-nature}

A novel attack vector on the \gls{qkd} algorithms BB84, its \ndimensional{4} variant, and \gls{ssp} using an
experimental research method will be investigated.
The attack vector analyzed involves a higher-dimensional measurement basis than that used by the legitimate parties.
The quantum simulator Cirq will be employed to model the \gls{qkd} algorithms and the proposed attack.
Doing so will produce quantitative data in the form of the error introduced by an eavesdropper and how much knowledge
said eavesdropper obtains when performing the high-dimensional eavesdropping attack.

%% file: sections/introduction/subsections/framework.tex
\subsection{Conceptual or Theoretical Framework} \label{subsec:introduction-framework}
% Include diagrams of Alice, Bob, and Eve
% Include diagrams of a quantum particle and what happens when it collapses.
% Include diagrams of how particles are measured  (Compare it to coin slots)
% Cite diagrams from other papers
% Outline variables of interest (dependent variables)
% Outline independent variables

Particles can be expressed as a matrix of probabilities that correspond to the likelihood of being measured in a
certain axis or state.
As per Heisenberg's uncertainty principle, the measurement of a particle can only be certain when its direction
coincides with one of the matrix's main axes.
When this happens, there will only ever be one state measured.
However, if the direction of a particle deviates from one of these main axes, their measurements will have a
probable error rate relative to the deviation.
Therefore, for each quantum variable, there exists a system of coordinates wherein the probable error ceases to exist
for that variable \citeHUP.

BB84 leverages this by using two encoding bases that exist in an even superposition between each other, causing a
particle in one encoding basis to have a deviation that results in a 50\% error rate relative to the other encoding
basis.
Due to this, when an eavesdropper measures a particle without knowing its encoding basis, this should result in each
intercepted particle having a 25\% chance of being measured incorrectly by the legitimate parties when in an
environment with no noise~\cite{Salas2024, Alhazmi2023}.
However, this paper speculates that if an eavesdropper were to represent the particles in a quantum channel whose
axes span all possible encoding states used, the error rate produced by the deviation from the main axes should be
reduced so long as each axis of the expanded matrix is equivalent to each possible direction a particle in the
quantum channel can be.

%\expandupon{This could be performed via sensors (A programmable qudit-based quantum processor (nature))}

% NOTE: May need to move the Qudits section in the Lit Review to here, with the transition of 'The idea of measuring particles in more than two dimensions is leveraged by...'

% Independent and dependent variables.
As mentioned above, the error in quantum particles measured, or \gls{qber}, can be used to detect an eavesdropper.
As such, the dependent variables of interest will be the \gls{qber} produced by an eavesdropper and how much knowledge
of the shared keys said eavesdropper obtains.
These variables are dependent on the data sent through the quantum channel and the number of encoding bases and/or
dimensions used in the \gls{qkd} algorithm.
The encoding bases and the dimension of the particles will be dictated by the \gls{qkd} algorithm used.

Three algorithms will be examined, BB84, a high-dimensional variant of BB84, and \gls{ssp}.
BB84 uses two encoding bases on a \ndimensional{2} particle~\cite{Bennett1984}, whereas its high-dimensional variant
uses two encoding bases on a \ndimensional{4} particle~\cite{BechmannPasquinucci2000}.
The final algorithm, \gls{ssp}, uses three encoding bases on a \ndimensional{2} particle~\cite{Bruss1998}.
These algorithms are further discussed in Sections~\ref{sec:lit-review} and~\ref{sec:methodology}.

%% file: sections/introduction/subsections/hypotheses.tex
\subsection{Hypotheses/Research Questions} \label{subsec:introduction-hypotheses}

This experimental study will test the following hypotheses:

\textbf{Hypothesis 1.} When an eavesdropper uses a high dimensional measurement basis to measure the quantum channel
being used to perform BB84 or its variants that rely on the same principle, the produced \gls{qber} will be minimal.
As such, an analysis of the \gls{qber} would not be a valid means of detecting this attack.

\textbf{Hypothesis 2.} Given Hypothesis 1, this can be mitigated by using a \gls{hdqkd} variant of BB84 and its
variants, or by using more encoding bases.
Doing so should increase the number of dimensions an eavesdropper must use ($d_e$) to Equation~\ref{eq:hypothesis},
wherein $d_{ab}$ is the dimension of the quantum particle used to transmit data and $b$ is the number of
encoding/decoding bases used by the legitimate parties.

\begin{equation} \label{eq:hypothesis}
    d_e = d_{ab} * b
\end{equation}

% Hypothesis 1:
% When an easdropper uses a high dimensional measurement basis to measure the quantum channel being used to perform BB84 or its variants, the produced QBERR will be minimal. As such, an analysis of the QBERR would not be a valid means of detection.

% Hypothesis 2:
% Given Hypothesis 1, this can be mitigated by using a HD-QKD variant of BB84 or by using more encoding bases. Doing so should increase the number of dimensions an eavesdropper must use to |[ x * n ]| wherein x is the dimension of the quantum particle used, and n is the number of encoding dimensions.

% Hypothesis 1:
% When an eavesdropper uses a 4th dimensional base to measure the quantum channel being used in BB84 or E91, the produced QBERR should be minimal. As such, an analysis of the QBERR would not be a valid means of detection.

% Hypothesis 2:
% Given Hypothesis 1, this can be prevented by using a HD-QKD variant of BB84 and E91 that uses the additional basis for security as opposed to noise reduction. <This may not actually be feasible though, especially since it would increase the key size since there would be a lower chance of guessing a correct base>.

% Hypothesis 3 (Maybe):
% When an eavesdropper uses an nth dimensional base to measure the quantum channel being used in a QKD algorithm, the produced QBERR should be minimal. This should hold true so long

%% file: sections/introduction/subsections/assumptions.tex
%\subsection{Assumptions}
%\subsection{Scope, Limitations, and Delimitations}

\subsection{Assumptions and Limitations} \label{subsec:introduction-assumption}

% https://phdstudent.com/thesis-and-dissertation-survival/research-design/stating-the-obvious-writing-assumptions-limitations-and-delimitations/
% http://researchdesign.lucalongo.eu/material/Assumptions-Limitations-Delimitations-and-Scope-of-the-Study.pdf
Without access to physical hardware, the quantum simulator Cirq was used to test the hypotheses.
Due to this, it must be assumed that a high-dimensional eavesdropping attack is physically possible.
Furthermore, the assumption that no noise was present also had to be made due to time constraints.
Cirq natively supports modeling noise on \ndimensional{2} quantum particles and provides the resources necessary to
define custom logic that models noise on quantum particles with a higher dimension.
However, doing so for complex experiments requires the definition of a custom \verb|NoiseModel|
subclass~\cite{CirqDevelopers2024}.
Creating this subclass for each test proved to be too time-consuming given a limited time frame.

%% file: sections/lit-review/lit-review.tex
\section{Literature Review} \label{sec:lit-review}

%https://fepbl.com/index.php/csitrj/article/view/815
% APA Usages (AES Usage.pdf, Atadoga2024)

\subimport{subsections}{history-of-qkd-algs.tex}

\subimport{subsections}{research-on-security-of-qkd.tex}
%\subimport{subsections}{quantum-cloning.tex}
\subimport{subsections}{qudits.tex}

%% file: sections/lit-review/subsections/history-of-qkd-algs.tex
\subsection{History of Quantum Key Distribution Algorithms} \label{subsec:lit-review-history}
%https://dl.acm.org/doi/pdf/10.1145/3575879.3576022
% Good source in general

\gls{qkd} protocols are key distribution algorithms that leverage quantum mechanics to secure and share secret keys
using a quantum channel~\cite{James2023}.
The first such protocol, BB84, was proposed in 1984 by Charles Bennett and Gilles Brassard~\cite{Sabani2022}.

When using BB84, the sender encodes classical data using photons that serve as quantum bits, or qubits.
This is accomplished by polarizing photons in one of two different bases before transmitting them through a
quantum channel.
The receiver then must guess which basis was used when measuring said photons~\cite{Singh2022}.

% E91
In 1991, Artur Ekert proposed the \gls{qkd} protocol E91.
This algorithm operates by sending pairs of entangled photons through a quantum channel, with entanglement being
defined as two or more particles that cannot be measured as individual independent states~\cite{Djordjevic2024}.
Entanglement is utilized by E91 to provided security as entangled particles should produce a \gls{chsh} correlation
value of $-2\sqrt{2}$ when undisturbed.
As such, a \gls{chsh} correlation value that differs from $-2\sqrt{2}$ indicates a lack of entanglement and the
presence of an eavesdropper~\cite{Ekert1991}.

% B92
BB84 was expanded upon by Charles Bennett in 1992, presenting a new \gls{qkd} algorithm.
This algorithm, B92, is a variant of BB84 that encodes classical data using two states as opposed to four.
Instead of an encoding basis being randomly selected for each bit, the value being encoded determines which basis
to use~\cite{Bennett1992}.

% SSP
Another \gls{qkd} algorithm, \gls{ssp}, was defined by~\citeauthor{Bruss1998} in~\citeyearNP{Bruss1998} and further
investigated by~\citeauthor{BechmannPasquinucci1999} in~\citeyearNP{BechmannPasquinucci1999}.
\Gls{ssp} is a variant of BB84 that uses a third encoding basis conjugate to the rectilinear and diagonal bases of BB84.
Doing so increases the \gls{qber} produced by an eavesdropper at the cost of an increase of discarded particles.
This is because two legitimate parties will only have a one in three chance of using the same encoding and decoding
basis~\cite{Bruss1998}.

Finally, in~\citeyearNP{Scarani2004}, a variant of BB84 was defined titled SARG04.
This algorithm was designed to be secure against \gls{pns} attacks.
The \gls{pns} attack exploits implementations of \gls{qkd} protocols that use weak light pulses, as light pulses
contain many photons. \unsure{many photons, make more precise.}
The attack operates by having an eavesdropper splitting off a single photon from the signal pulse, enabling the
measurement of the quantum channel without the introduction of noise~\cite{Sabani2022}.
SARG04 protects against this attack by altering how parties determine which encoding bases were used for each
qubit.
Instead of announcing the basis used on a classical channel, a pair of non-orthogonal states are sent via the quantum
channel.
This requires an eavesdropper to split off more photons from the quantum channel when performing a \gls{pns} attack,
making their detection easier~\cite{Scarani2004}.

%% file: sections/lit-review/subsections/research-on-security-of-qkd.tex
\subsection{Research on the Security of QKD} \label{subsec:lit-review-research-on-security}

\subsubsection{MitM Attacks on BB84}

\begin{figure}[h]
    \caption{A depiction of a \gls{qkd} setup both with and without an eavesdropper.}
    \label{fig:lit-review-mitm-steps}
    \centering
    \includegraphics[width=11.5cm]{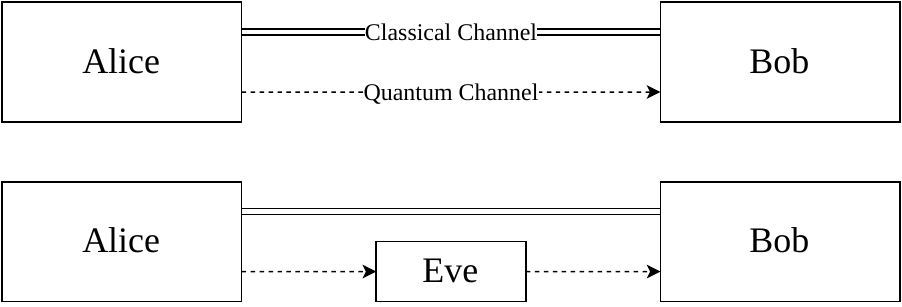}
\end{figure}

Most research on the security of \gls{qkd} algorithms has been centered on BB84's ability to detect an eavesdropper
performing a \gls{mitm}, or intercept-resend, attack.
Such attacks aim to obtain the encryption key being shared without the knowledge of the intended
recipients~\cite{Jawad2023}.
This can be achieved by cutting the quantum channel between Alice and Bob before connecting the severed ends to a
pair of \gls{qkd} devices capable of measuring and encoding photons~\cite{Alhazmi2023, Liliana2019}.
This process is demonstrated in Figure~\ref{fig:lit-review-mitm-steps}.

To measure the intercepted photons, Eve must follow the same steps as Bob, selecting a random decoding basis for each
photon before measurement.
Eve must then encode their measurements in its respective basis before sending them to Bob~\cite{Huang2009}.

\subsubsection{Detection of Eve}

If Eve randomly chooses the same base as Alice, the state of the electron will not be altered.
However, if Eve measures the particle using a different base, the particle will randomly collapse into either a 1 or 0
with equal probability.
Because of this, Bob has a 25\% chance of measuring each intercepted photon incorrectly when in an environment with no
noise.
As such, the probability of Eve being detected $P_d$ given $K$ qubits are given by the formula
$P_d=1-(3/4)^K$~\cite{Salas2024, Alhazmi2023}.

% \topic{Min Qubits for Security With \& Without No Noise}

%% file: sections/lit-review/subsections/qudits.tex
\subsection{High-Dimensional Quantum Key Distribution} \label{subsec:lit-review-qudits}

Qubits only use two states of a physical property to store and process information.
However, many such properties have multiple usable states.
This can include properties such as particle frequency, energy level, spin state~\cite{Wang2020} and
\gls{oam}~\cite{Zi2023}.
Quantum bits that use these additional dimensions are called qudits.
This can be leveraged by \gls{hdqkd} to encode data as qudits as opposed to qubits, often increasing noise
resilience~\cite{Mueller2023}.
\citeauthor{BechmannPasquinucci2000} demonstrated how qudits could be utilized in \gls{qkd} algorithms by defining a
\ndimensional{4} variant of BB84 in~\citeyearNP{BechmannPasquinucci2000}.
This was achieved by redefining the encoding bases used by BB84 to account for the use of \ndimensional{4} qudits, with
each qudit representing two bits.
Most research on \gls{hdqkd} algorithms have analyzed high-dimensional variants of BB84, showing that \gls{hdqkd}
results in greater key efficiency as well as an increase in \gls{qber} when an eavesdropper is
present~\cite{Liliana2019, Zahidy2024}.

% Quantum Cloning
An example of the security offered by \gls{hdqkd} can be seen in their resistance to quantum cloning attacks.
\gls{qkd} algorithms rely on the no-cloning theorem to provide security~\cite{Sabani2022}, which states that it is
impossible for one to perfectly copy a quantum state without altering it~\cite{Wootters1982}.
Despite this, the creation of imperfect copies of quantum states is possible using quantum cloning
machines~\cite{Idan2024, Iqbal2023}.

\citeA{Bouchard2017} explored the use of such a machine to perform a \gls{mitm} attack on
BB84 in a paper titled ``High-dimensional quantum cloning and applications to quantum hacking''.
It detailed a high-dimensional \gls{uqcm} capable of creating imperfect clones of the \gls{oam} states of photons.
This was achieved using two input qubits, the first being the qubit sent between Alice and Bob, and the
second being a qubit in an even superposition between all $d$ dimensions of the first qubit.
By passing both photons through a 50:50 beam splitter, they will exit said beam splitter in a similar, if not identical,
state due to their bosonic nature.
The \gls{uqcm} was used to attack both high-dimensional and standard versions of BB84, finding that the
high-dimensional protocol made an eavesdropper's presence clearly visible and demonstrating the improved security of
\gls{hdqkd} compared to their lower dimensional variants.

%% file: sections/methodology/methodology.tex
%    NOTE: Make sure to define Alice, Bob, and Eve here.
%    NOTE: Define expected control errors in results section.
%    TODO: Detail how gates can be represented as matrices.

%    Describe process used for each algorithm.
%    Step 1: Gate creation for normal algorithm.
%    Step 2: Create and verify a control.
%    Step 3: Gate creation for scenarios (detail qudit limitation in cirq). [also define states of qudits in this section]
%    Step 4: Tested two scenarios wherein an eavesdropper was present.
%       Scenario 1....
%       Scenario 2....
%       The results of both scenarios should be identical.
%    Step 5: Run simulations with control, scenario 1, and scenario 2 [state that measurements can be seen in measurement appendix]
%
%    Each subfile elaborates on each step in further detail (with the exception of step 5,
%    which is elaborated in the results section).

\section{Methodology} \label{sec:methodology}

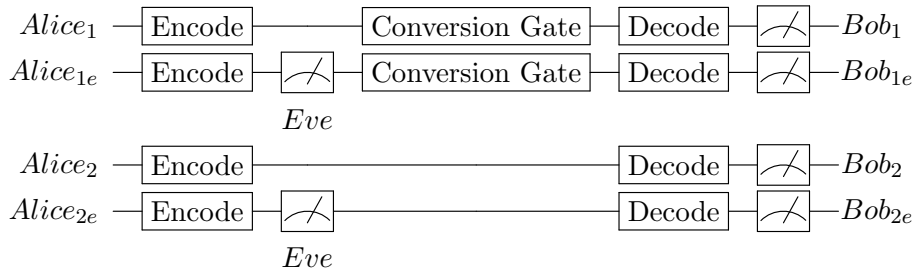
\begin{figure}[h!]
    \centering
    \caption{The gates used in \scenarioA~(denoted by $1$) and \scenarioB~(denoted by $2$).}
    \label{fig:methodology-scenarios}
    \scalebox{1.0}{
        \Qcircuit @C=1.0em @R=0.2em @!R { \\
        & \lstick{Alice_{1}}  & \gate{\mathrm{Encode}} & \qw    & \gate{\mathrm{Conversion\ Gate}} & \gate{\mathrm{Decode}} & \meter & \qw & \ Bob_{1}    \\
        & \lstick{Alice_{1e}} & \gate{\mathrm{Encode}} & \meter & \gate{\mathrm{Conversion\ Gate}} & \gate{\mathrm{Decode}} & \meter & \qw & \ \ Bob_{1e} \\
        &                     &                        & Eve \\
        & \lstick{Alice_{2}}  & \gate{\mathrm{Encode}} & \qw    & \qw & \gate{\mathrm{Decode}} & \meter & \qw & \ Bob_{2}    \\
        & \lstick{Alice_{2e}} & \gate{\mathrm{Encode}} & \meter & \qw & \gate{\mathrm{Decode}} & \meter & \qw & \ \ Bob_{2e} \\
        &                     &                        & Eve \\
        \\ }}
\end{figure}

Key exchanges were performed between two parties, Alice and Bob.
When present, the eavesdropper is referred to as Eve.
An unmodified version, or control group, of each \gls{qkd} algorithm was analyzed alongside the impacts of Eve when
performing a \gls{mitm} attack.
In addition to the control group, two different means of simulating high-dimensional eavesdropping, \scenarioA~and
\scenarioB, were tested.
This is because Cirq cannot measure qudits in a dimension different from said qudits
dimensions~\cite{CirqDevelopers2024}.
To account for this, \scenarioA~converts the high-dimensional qudit encoded by Alice to a lower-dimensional qudit
whose dimensions are equal to that of the algorithm's control group.
\scenarioB~circumvented this limitation by having Bob measure the high-dimensional qudit and altering the
resulting binary values to reflect their expected values given the state measured.
Figure~\ref{fig:methodology-scenarios} depicts the logic for \scenarioA~and \scenarioB.

Cirq measures qudits in a range of $0$ to $d-1$, where $d$ is the dimensions of the qudit.
These states should correlate to a binary value and its corresponding encoding basis.
For example, the qudit states when testing BB84 correlated to the values in
Equation~\ref{eq:methodology-qudit-measurement}.
\begin{equation} \label{eq:methodology-qudit-measurement}
    \begin{split}
        |0\rangle = \text{0 encoded in the rectilinear basis} \\
        |1\rangle = \text{0 encoded in the diagonal basis}    \\
        |2\rangle = \text{1 encoded in the rectilinear basis} \\
        |3\rangle = \text{1 encoded in the diagonal basis}
    \end{split}
\end{equation}
Given these states, Bob can replace measurements of two with one and change measurements of one and three to a random
bit value.
Doing so models the expected behavior of a qubit when represented with a \ndimensional{4} qudit.
These alterations, alongside the corresponding binary value of the states measured by Eve, can be seen
in Appendix~\ref{app:res-format}.

%Two means of simulation were performed because a measurement gate in Cirq cannot measure qudits in a dimension
%different from said qudits dimensions~\cite{CirqDevelopers2024}.
%Due to this, Alice had to encode the quantum channel using qudits whose dimensions equal that used by Eve.
%To account for this, \scenarioA converts the high-dimensional qudit encoded by Alice to a lower-dimensional qudit
%whose dimensions are equal to that of the algorithm's control group.
%\scenarioB circumvented this limitation by having Bob measure the high-dimensional qudit and altering the
%resulting binary values to reflect their expected values given the state measured.
%These alterations, alongside the corresponding binary value of the states measured by Eve, can be seen
%in~\ref{app:res-format}.
%The results produced by these different implementations of high-dimensional eavesdropping should be identical.

The high-dimensional gate variants were prepended with ``Qid'' and defined using matrices.
The states of a $d$-dimensional qudit can be represented by a set of orthonormal state vectors as
shown in Equation~\ref{eq:methodology-qudits}, wherein $\alpha_d$ is the probability of measuring $|d\rangle$ and
$|\alpha_0|^2 + |\alpha_1|^2 \ldots + |\alpha_{d-1}|^2 = 1$~\cite{Wang2020}.
\begin{equation} \label{eq:methodology-qudits}
    |\alpha\rangle = \alpha_0|0\rangle + \alpha_1|1\rangle + \ldots + \alpha_{d-1}|d-1\rangle =
    \begin{pmatrix}
        \alpha_0 \\
        \alpha_1 \\
        \vdots   \\
        \alpha_{d-1}
    \end{pmatrix}
\end{equation}
Quantum gates can also be represented as a matrix whose values correspond to the probability of measuring a given
state given an input vector~\cite{Spranger2023}.\todo{as can be seen in...}

\insertobject{Draw a matrix whose top is labeled with ``Input States <newline> |0> and |1>''. Label row 1 as chance to measure |0> ($|value|^2$) given a specific input state and label row 2 in the same manner.}
% https://tex.stackexchange.com/questions/59517/label-rows-of-a-matrix-by-characters

When applying a quantum gate to $n$ qudits, one must first calculate the joint states of said
qudits~\cite{Spranger2023}.
Consequently, applying a gate to multiple qudits with different dimensions will produce a matrix whose size is the sum
of the squares of each qudit's dimensions.
For example, a quantum gate that operates on an \ndimensional{8} qudit, a \ndimensional{2} qubit, and a
\ndimensional{4} qudit will result in a matrix of size $64^2$.

\subimport{subsections}{bb84.tex}

\subimport{subsections}{hdbb84.tex}

\subimport{subsections}{ssp.tex}

%% file: sections/methodology/subsections/bb84.tex
\subsection{BB84} \label{subsec:methodology-bb84}

%  Step 2: Create and verify control (no custom gates where needed to analyze BB84)
\subsubsection{Control Group}

BB84 relies on two conjugate bases, the rectilinear and diagonal basis.
The use of conjugate bases ensures that when a particle is in a specific state of one basis,
measuring said particle in the other basis results in a random output.
The rectilinear basis is defined as a particle with a polarization of either 0\degree~or
90\degree, whereas the diagonal basis is defined as a particle with a polarization of either
45\degree~or 135\degree~\cite{Bennett1984}.

To perform a key exchange via BB84, Alice must first generate a random key that will be encoded and shared with Bob
and choose an encoding basis for each bit of the key~\cite{Bennett1984}.
A \gls{hgate} can be used to encode or decode a qubit in the diagonal basis.
The properties of the \gls{hgate} are provided in Equation~\ref{eq:hgate-sv}.
An X-gate is applied to a qubit before transmission if a binary 1 is being encoded.
To decode the quantum channel, Bob must choose a random measurement basis for each qubit received.
When present, Eve follows the same procedure as Bob~\cite{Saeed2022}.

\begin{equation} \label{eq:hgate-sv}
    \begin{split}
        H|0\rangle = \frac{1}{\sqrt{2}}(|0\rangle + |1\rangle) = |+\rangle \\
        H|1\rangle = \frac{1}{\sqrt{2}}(|0\rangle - |1\rangle) = |-\rangle
    \end{split}
\end{equation}

Once Bob has received and measured all qubits sent by Alice, the two generate a sifted key.
This is accomplished by sharing the encoding and decoding basis used for each qubit, and discarding
bits measured in different bases~\cite{Bennett1984}.
The two parties then check for the presence of an eavesdropper by comparing a number of bits in the sifted key.
Without an eavesdropper or noise, all compared bits of the sifted key should be identical~\cite{Bennett1984, Saeed2022}.

% Step 3: Gate creation for eavesdropper.
\subsubsection{Gate Creation for HD-Eavesdropping}

% Qudit state values.
The states of Eve's qudit are given in Equation~\ref{eq:bb84-eve-qudit}.
Since Cirq is unable to measure qudits using a different dimension, high-dimensional
variants of the X- and H-gate had to be defined that would change the state of a qudit
based on how their qubit equivalent would normally affect a qubit.
In addition to this, \scenarioA~required the creation of several additional gates used to
convert a \ndimensional{4} qudit to a single qubit.

\begin{equation} \label{eq:bb84-eve-qudit}
    \begin{split}
        |0\rangle = |0\rangle \\
        |1\rangle = |+\rangle \\
        |2\rangle = |1\rangle \\
        |3\rangle = |-\rangle
    \end{split}
\end{equation}

% The gates below will be a paragraph that starts with the name of the gate in bold.
% Qid-X Gate.
\paragraph{\QidXBB.} The X-gate is the equivalent of a classical bit flip, and has the properties outlined in
Equation~\ref{eq:xgate}~\cite{Nielsen2010}.
A high-dimensional variant of the X-gate that reflects this is provided in Equation~\ref{eq:qidx-1-properties}.

\begin{equation} \label{eq:xgate}
    X\begin{pmatrix} \alpha \\ \beta \end{pmatrix} = \begin{pmatrix} \beta \\ \alpha \end{pmatrix}
\end{equation}

\begin{equation} \label{eq:qidx-1-properties}
    \begin{split}
        X|0\rangle = |2\rangle \\  % X was \sigma_x
        X|1\rangle = |1\rangle \\
        X|2\rangle = |0\rangle \\
        X|3\rangle = -|3\rangle
    \end{split}
\end{equation}

% Qid-H Gate.
\paragraph{\QidHBB.} Equation~\ref{eq:hgate} lists the properties of the \gls{hgate}~\cite{Qiskit2023}.
A high-dimensional variant of the \gls{hgate} that reflects this is provided in Equation~\ref{eq:qidh-1-properties}.

\begin{equation} \label{eq:hgate}
    \begin{split}
        H|0\rangle = |+\rangle \\
        H|1\rangle = |-\rangle \\
        H|+\rangle = |0\rangle \\
        H|-\rangle = |1\rangle
    \end{split}
\end{equation}

\begin{equation} \label{eq:qidh-1-properties}
    \begin{split}
        H|0\rangle = |1\rangle \\
        H|1\rangle = |0\rangle \\
        H|2\rangle = |3\rangle \\
        H|3\rangle = |2\rangle
    \end{split}
\end{equation}

% \expandupon{It should be noted that the QidH gate does not create an even superposition (maybe)}

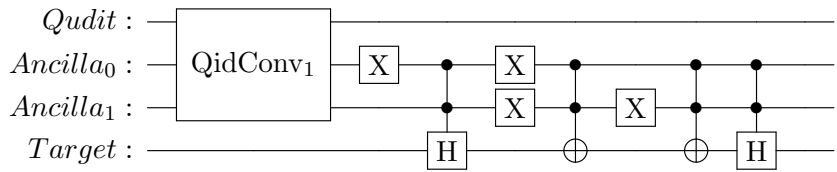
\begin{figure}[h!]
    \centering
    \caption{BB84 conversion gate.}
    \label{fig:bb84-conv-circ}
    \scalebox{1.0}{
        \Qcircuit @C=1.0em @R=0.2em @!R { \\
        \nghost{Qudit :  } & \lstick{Qudit :  } & \multigate{2}{\mathrm{QidConv_1}} & \qw & \qw & \qw & \qw & \qw & \qw & \qw & \qw & \qw\\
        \nghost{Ancilla_0 :  } & \lstick{Ancilla_0 :  } & \ghost{\mathrm{QidConv_1}} & \gate{\mathrm{X}} & \ctrl{1} & \gate{\mathrm{X}} & \ctrl{1} & \qw & \ctrl{1} & \ctrl{1} & \qw & \qw\\
        \nghost{Ancilla_1 :  } & \lstick{Ancilla_1 :  } & \ghost{\mathrm{QidConv_1}} & \qw & \ctrl{1} & \gate{\mathrm{X}} & \ctrl{1} & \gate{\mathrm{X}} & \ctrl{1} & \ctrl{1} & \qw & \qw\\
        \nghost{Target :  } & \lstick{Target :  } & \qw & \qw & \gate{\mathrm{H}} & \qw & \targ & \qw & \targ & \gate{\mathrm{H}} & \qw & \qw\\
        \\ }}
\end{figure}

% QidBB84Conv Gate.
\paragraph{BB84 Conversion Gate.} A conversion gate for BB84 was used in \scenarioA~to convert a \ndimensional{4} qudit
to a \ndimensional{2} qubit.
Figure~\ref{fig:bb84-conv-circ} depicts the logic for this gate, wherein \QidConvBB~is a unitary that maps a qudit to
two qubits that store the encoding basis and encoded value.
The matrix for this gate can be seen in Appendix~\ref{app:conv-unitaries}.

%% file: sections/methodology/subsections/hdbb84.tex
\subsection{HD-BB84} \label{subsec:methodology-hdbb84}

% Step 2: Create and verify control.
\subsubsection{Control Group}

In~\citeyearNP{BechmannPasquinucci2000}, \citeauthor{BechmannPasquinucci2000} proposed a \ndimensional{4} variant of
BB84 that uses two encoding bases on \ndimensional{4} qudits.
They defined the first basis, the \psibasis, as an arbitrarily chosen set of values given in
Equation~\ref{eq:hdbb84-psi-basis}, alongside their corresponding binary value.
\begin{equation} \label{eq:hdbb84-psi-basis}
    \begin{split}
        |\psi_\alpha\rangle = 00 \\
        |\psi_\beta\rangle = 01 \\
        |\psi_\gamma\rangle = 10 \\
        |\psi_\delta\rangle = 11
    \end{split}
\end{equation}
To create a symmetric protocol, the second ($\phi$) basis, had to satisfy the criteria outlined in
Equation~\ref{eq:hdbb84-phi-criteria}.
\begin{equation} \label{eq:hdbb84-phi-criteria}
    \begin{split}
        |\langle\psi_i|\psi_j\rangle| = \phi_{ij} \\
        |\langle\psi_i|\phi_j\rangle| = 1/2
    \end{split}
\end{equation}
Multiple bases exist which fulfill these requirements, but \citeauthor{BechmannPasquinucci2000} opted to use the
basis Equation~\ref{eq:hdbb84-phi-basis}.
\begin{equation} \label{eq:hdbb84-phi-basis}
    \begin{split}
        |\phi_\alpha\rangle = \frac{1}{2}(
            |\psi_\alpha\rangle + |\psi_\beta\rangle + |\psi_\gamma\rangle + |\psi_\delta\rangle
        ) \\
        |\phi_\beta\rangle = \frac{1}{2}(
            |\psi_\alpha\rangle - |\psi_\beta\rangle + |\psi_\gamma\rangle - |\psi_\delta\rangle
        ) \\
        |\phi_\gamma\rangle = \frac{1}{2}(
            |\psi_\alpha\rangle - |\psi_\beta\rangle - |\psi_\gamma\rangle + |\psi_\delta\rangle
        ) \\
        |\phi_\delta\rangle = \frac{1}{2}(
            |\psi_\alpha\rangle + |\psi_\beta\rangle - |\psi_\gamma\rangle - |\psi_\delta\rangle
        )
    \end{split}
\end{equation}

However, when trying to use this \phibasis, the results differed from the expected output.
Equation~\ref{eq:hdbb84-phi-basis-error} details the behavior that occurred when both Alice and Bob used the
\phibasis~to encode and decode the quantum channel.
\begin{equation} \label{eq:hdbb84-phi-basis-error}
    \begin{split}
        \phi^2|\psi_\alpha\rangle = |\psi_\alpha\rangle \\
        \phi^2|\psi_\beta\rangle = |\psi_\delta\rangle \\
        \phi^2|\psi_\gamma\rangle = |\psi_\beta\rangle \\
        \phi^2|\psi_\delta\rangle = |\psi_\gamma\rangle
    \end{split}
\end{equation}
Such behavior meant that Bob would measure a different value than that sent by Alice even when the two used the
\phibasis.
As such, the \phibasis~was redefined to be that in Equation~\ref{eq:hdbb84-phi-basis-fixed}.
\begin{equation} \label{eq:hdbb84-phi-basis-fixed}
    \begin{split}
        |\phi_\alpha\rangle = \frac{1}{2}(
            |\psi_\alpha\rangle + |\psi_\beta\rangle + |\psi_\gamma\rangle + |\psi_\delta\rangle
        ) \\
        |\phi_\beta\rangle = \frac{1}{2}(
            |\psi_\alpha\rangle + |\psi_\beta\rangle - |\psi_\gamma\rangle - |\psi_\delta\rangle
        ) \\
        |\phi_\gamma\rangle = \frac{1}{2}(
            |\psi_\alpha\rangle - |\psi_\beta\rangle + |\psi_\gamma\rangle - |\psi_\delta\rangle
        ) \\
        |\phi_\delta\rangle = \frac{1}{2}(
            |\psi_\alpha\rangle - |\psi_\beta\rangle - |\psi_\gamma\rangle + |\psi_\delta\rangle
        )
    \end{split}
\end{equation}

\expandupon[inline]{The phi basis uses 1/2 bc it is the sqrt of 1/4, find src for qntm prob matrix.}

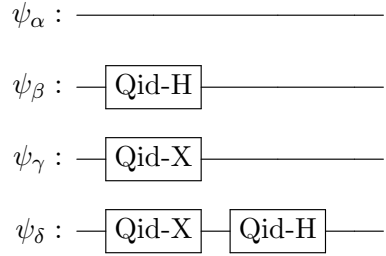
\begin{figure}[h!]
    \centering
    \caption{$\psi$ Gates}
    \label{fig:hdbb84-psi-gates}
    \scalebox{1.0}{
        \Qcircuit @C=1.0em @R=1.0em @!R { \\
        \nghost{{\psi}_{\alpha} :  } & \lstick{{\psi}_{\alpha} :  } & \qw & \qw & \qw & \qw\\
        \nghost{{\psi}_{\beta} :  } & \lstick{{\psi}_{\beta} :  } & \gate{\mathrm{Qid\mbox{-}H}} & \qw & \qw & \qw\\
        \nghost{{\psi}_{\gamma} :  } & \lstick{{\psi}_{\gamma} :  } & \gate{\mathrm{Qid\mbox{-}X}} & \qw & \qw & \qw\\
        \nghost{{\psi}_{\delta} :  } & \lstick{{\psi}_{\delta} :  } & \gate{\mathrm{Qid\mbox{-}X}} & \gate{\mathrm{Qid\mbox{-}H}} & \qw & \qw\\
        \\ }}
\end{figure}

Similar to BB84, the high-dimensional variant created to model high-dimensional eavesdropping entails Alice encoding a
photon before applying the randomly chosen encoding basis.
As shown in Equation~\ref{eq:hdbb84-psi-basis}, there are four states in each encoding basis wherein each state
represents two bits instead of one bit.
Consequently, half as many particles is needed compared to BB84.
Same as in BB84, Alice must first encode a quantum particle in the \psibasis~before applying the \phibasis.
To account for this, an $\alpha$-, $\beta$-, $\gamma$-, and $\delta$-gate were defined using the quantum gates required
to perform high-dimensional eavesdropping on BB84.
These gates are illustrated in Figure~\ref{fig:hdbb84-psi-gates}.

The process of performing a \gls{mitm} attack are identical to that in BB84, with the only difference being the usage
of the $\psi$- and $\phi$-basis instead of the rectilinear and diagonal basis.
It should also be noted that, when creating the sifted key, bits are discarded in two-bit chunks.
In other words, two bits are either kept or discarded depending on whether Alice and Bob used the same or different
encoding bases respectively~\cite{BechmannPasquinucci2000}.

% Step 3: Gate creation for eavesdropper.
\subsubsection{Gate Creation for HD-Eavesdropping}

% Qudit state values.
The states of Eve's qudit when performing the high-dimensional eavesdropping attack are given in
Equation~\ref{eq:hdbb84-eve-qudit}.
Once again, high-dimensional variants of the gates used to perform this high-dimensional BB84 variant had to be
defined alongside gates to convert the \ndimensional{8} qudit to a \ndimensional{4} qudit.

\begin{equation} \label{eq:hdbb84-eve-qudit}
    \begin{split}
        |0\rangle = |\psi_\alpha\rangle \\
        |1\rangle = |\phi_\alpha\rangle \\
        |2\rangle = |\psi_\beta\rangle  \\
        |3\rangle = |\phi_\beta\rangle  \\
        |4\rangle = |\psi_\gamma\rangle \\
        |5\rangle = |\phi_\gamma\rangle \\
        |6\rangle = |\psi_\delta\rangle \\
        |7\rangle = |\phi_\delta\rangle
    \end{split}
\end{equation}

\paragraph{Qid-$\psi$ and Qid-$\phi$ Gates.} Table~\ref{tab:hdbb84-qid-gates} depicts the high-dimensional variants of the
gates used to manipulate qudits in the $\psi$ basis, alongside the high-dimensional variant of the $\phi$-gate.

\begin{table}[h!]
    \centering
    \caption{The results of a given input state to their corresponding output state when a high-dimensional variant of
    the $\alpha$-, $\beta$-, $\gamma$-, $\delta$-, or $\phi$-gate is applied.}
    \label{tab:hdbb84-qid-gates}
    \begin{tabular}{c c c c c c}
        Input State & Qid-$\alpha$ & Qid-$\beta$ & Qid-$\gamma$ & Qid-$\delta$ & Qid-$\phi$  \\ \hline
        $|0\rangle$ & $|0\rangle$  & $|2\rangle$ & $|4\rangle$  & $|6\rangle$  & $|1\rangle$ \\
        $|1\rangle$ & $|1\rangle$  & $|3\rangle$ & $|5\rangle$  & $|7\rangle$  & $|0\rangle$ \\
        $|2\rangle$ & $|2\rangle$  & $|0\rangle$ & $|6\rangle$  & $|4\rangle$  & $|3\rangle$ \\
        $|3\rangle$ & $|3\rangle$  & $|1\rangle$ & $|7\rangle$  & $|5\rangle$  & $|2\rangle$ \\
        $|4\rangle$ & $|4\rangle$  & $|6\rangle$ & $|0\rangle$  & $|2\rangle$  & $|5\rangle$ \\
        $|5\rangle$ & $|5\rangle$  & $|7\rangle$ & $|1\rangle$  & $|3\rangle$  & $|4\rangle$ \\
        $|6\rangle$ & $|6\rangle$  & $|4\rangle$ & $|2\rangle$  & $|0\rangle$  & $|7\rangle$ \\
        $|7\rangle$ & $|7\rangle$  & $|5\rangle$ & $|3\rangle$  & $|1\rangle$  & $|6\rangle$
    \end{tabular}
\end{table}

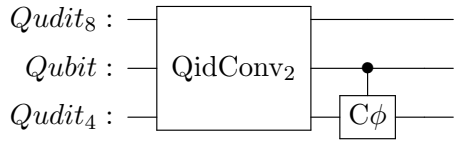
\begin{figure}[h!]
    \centering
    \caption{HD-BB84 conversion gate.}
    \label{fig:hdbb84-conv-circ}
    \scalebox{1.0}{
        \Qcircuit @C=1.0em @R=0.2em @!R { \\
        \nghost{Qudit_8 :  } & \lstick{Qudit_8 :  } & \multigate{2}{\mathrm{QidConv_2}} & \qw & \qw & \qw\\
        \nghost{Qubit :  } & \lstick{Qubit :  } & \ghost{\mathrm{QidConv_2}} & \ctrl{1} & \qw & \qw\\
        \nghost{Qudit_4 :  } & \lstick{Qudit_4 :  } & \ghost{\mathrm{QidConv_2}} & \gate{\mathrm{C\phi}} & \qw & \qw\\
        \\ }}
\end{figure}

% QidHDBB84Conv Gate.
\paragraph{HD-BB84 Conversion Gate.} A conversion gate for the \ndimensional{4} variant of BB84 was used in
\scenarioA~to convert an \ndimensional{8} qudit to a \ndimensional{4} qudit.
The logic for this gate can be seen in Figure~\ref{fig:hdbb84-conv-circ}.
The \QidConvHDBB~gate maps the states of an \ndimensional{8} qudit to a \ndimensional{4} qudit and a \ndimensional{2}
ancilla qubit.
The qubit is used to determine what encoding basis to apply using a conditional $\phi$-gate (C$\phi$), while the
\ndimensional{4} qudit will be the value measured by Bob.
The unitary for \QidConvHDBB~can be seen in Appendix~\ref{app:conv-unitaries}.

%% file: sections/methodology/subsections/ssp.tex
\subsection{SSP} \label{subsec:methodology-ssp}

% Step 2: Create and verify control.
\subsubsection{Control Group}

The two axes BB84 uses to encode qubits are the z- and x-axis for the rectilinear and diagonal basis
respectively.
However, \citeauthor{Bennett1984} provided another possible basis conjugate to both the rectilinear and diagonal
bases that could be used instead.
This basis exists on the y-axis and consists of the circular
polarizations Equation~\ref{eq:ssp-circular-polarization-bennett}~\cite{Bennett1984}.

\begin{equation} \label{eq:ssp-circular-polarization-bennett}
    |i\rangle = \frac{1}{\sqrt{2}}
    \begin{pmatrix}
        1 \\
        i
    \end{pmatrix}
    \qquad  % Insert space between matrices.
    |-i\rangle = \frac{1}{\sqrt{2}}
    \begin{pmatrix}
        i \\
        1
    \end{pmatrix}
\end{equation}

\citeauthor{Bruss1998} expanded upon this in~\citeyearNP{Bruss1998} with \gls{ssp}, a variant of BB84 that uses
the x-, y-, and z-axes, and was further investigated by~\citeauthor{BechmannPasquinucci1999}
in~\citeyearNP{BechmannPasquinucci1999}.
Though the y-axis for \gls{ssp} was redefined to be that in Equation~\ref{eq:ssp-circular-polarization-bruss}.

\begin{equation} \label{eq:ssp-circular-polarization-bruss}
    |i\rangle = \frac{1}{\sqrt{2}}
    \begin{pmatrix}
        1 \\
        i
    \end{pmatrix}
    \qquad  % Insert space between matrices.
    |-i\rangle = \frac{1}{\sqrt{2}}
    \begin{pmatrix}
        1 \\
        -i
    \end{pmatrix}
\end{equation}

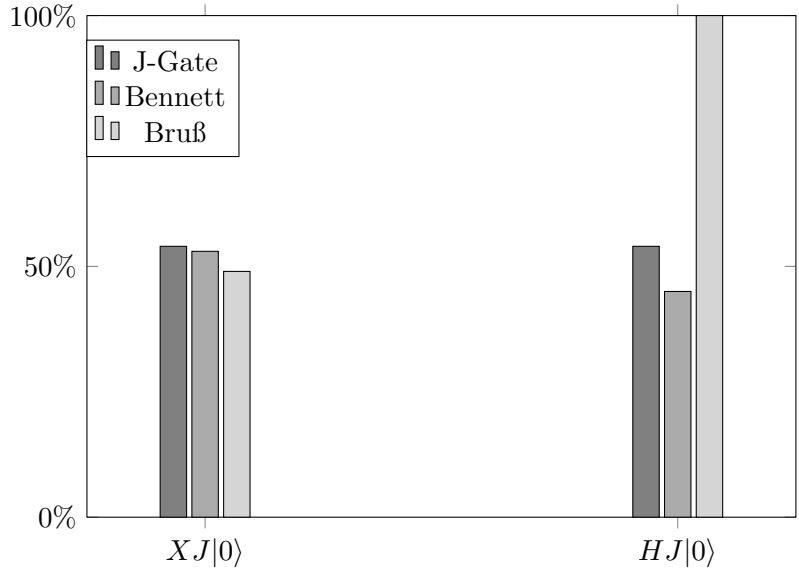
\begin{figure}
    \caption{The percentage of $0$'s measured after an X-gate or H-gate has been applied to 100 qubits encoded in
    the y-axis given the states defined in this paper, \protect\citeauthor{Bennett1984},
    and~\protect\citeauthor{Bruss1998}.}
    \label{fig:ssp-proposed-yaxes}
    \begin{center}
        \begin{tikzpicture}
            \begin{axis} [
            ybar,
            ymin=0,
            ymax=1,
            ytick={0.0, 0.5, 1.0},
            yticklabels={0\%, 50\%, 100\%},
            symbolic x coords={$XJ|0\rangle$, $HJ|0\rangle$},
            xtick={$XJ|0\rangle$, $HJ|0\rangle$},
            enlarge x limits = 0.25,
            width=\textwidth/1.5,
            height=\textwidth/2,
            legend style={at={(-0.001, 0.835)}, anchor=west},
            ]
            \addplot [nthBarOfThree=1] table [x=gates, y=J-Gate]{\proposedYaxesData};
            \addplot [nthBarOfThree=2] table[x=gates, y=Bennett]{\proposedYaxesData};
            \addplot [nthBarOfThree=3] table[x=gates, y=Bruss]{\proposedYaxesData};
            \legend{J-Gate, Bennett, \Bruss}
            \end{axis}
        \end{tikzpicture}
    \end{center}
\end{figure}

However, neither basis could be used.
For a basis to be secure, measurement on a different basis should result in a random value.
As such, when an X- or H-gate is applied to a particle encoded in the y-axis, the results measured should be random.
As can be seen in Figure~\ref{fig:ssp-proposed-yaxes}, the states defined by~\citeauthor{Bennett1984} meet said
criterion, whereas the states provided by~\citeauthor{Bruss1998} do not.

% Data for bar chart.
\pgfplotstableread[row sep=\\,col sep=&]{
    gates         & J-Gate & Bennett & Bruss \\
    $XJ|0\rangle$ & 0.54   & 0.53    & 0.49  \\
    $HJ|0\rangle$ & 0.54   & 0.45    & 1.0   \\
}\proposedYaxesData

Another criterion that must be met is that applying the basis twice should result in the original value when measured.
As can be seen in Equation~\ref{eq:ssp-bennett-yaxis-twice}, the states provided by~\citeauthor{Bennett1984} do not
achieve this.
When applying their basis twice, it would result in the inverse of the encoded value.

\begin{equation} \label{eq:ssp-bennett-yaxis-twice}
    \frac{1}{\sqrt{2}}\begin{pmatrix}
        1 & i \\
        i & 1
    \end{pmatrix}^2
    =
    \begin{pmatrix}
        0 & i \\
        i & 0
    \end{pmatrix}
\end{equation}

As such, the circular polarization had to be redefined to Equation~\ref{eq:ssp-circular-polarization-jgate}.
The corresponding gate to apply this basis was defined as the J-gate\footnote{Y is commonly associated with the
Pauli-Y gate and I is commonly associated with the Identity gate~\cite{Qiskit2023}. As such, J was selected
because $j$ is a commonly used alternative for $i$~\cite{Du2023}.}.
As can be seen in Figure~\ref{fig:ssp-proposed-yaxes} and Equation~\ref{eq:ssp-jgate-twice}, this basis satisfies both
criteria.

\begin{equation} \label{eq:ssp-circular-polarization-jgate}
    |i\rangle = \frac{1}{\sqrt{2}}
    \begin{pmatrix}
        i \\
        1
    \end{pmatrix}
    \qquad  % Insert space between matrices.
    |-i\rangle = \frac{1}{\sqrt{2}}
    \begin{pmatrix}
        -1 \\
        -i
    \end{pmatrix}
\end{equation}

\begin{equation} \label{eq:ssp-jgate-twice}
    \frac{1}{\sqrt{2}}\begin{pmatrix}
        i & -1 \\
        1 & -i
    \end{pmatrix}^2
    =
    \begin{pmatrix}
        -1 & 0 \\
        0 & -1
    \end{pmatrix}
\end{equation}

% Step 3: Gate creation for eavesdropper.
\subsubsection{Gate Creation for HD-Eavesdropping}

% Qudit state values.
The states of Eve's qudits are given in Equation~\ref{eq:ssp-eve-qudit}.
Same as with BB84 and its high-dimensional variant, high-dimensional X-, H-, and J-gates had to be defined in addition
to gates that convert a \ndimensional{6} qudit to a qubit.

\begin{equation} \label{eq:ssp-eve-qudit}
    \begin{split}
        |0\rangle = |0\rangle \\
        |1\rangle = |+\rangle \\
        |2\rangle = |i\rangle \\
        |3\rangle = |1\rangle \\
        |4\rangle = |-\rangle \\
        |5\rangle = |-i\rangle
    \end{split}
\end{equation}

\paragraph{Qid-X$_{2}$-, H$_{2}$-, and J-Gates.} The high-dimensional variants of the X- and H-gates were obtained by
modifying those created when testing BB84.
The results of applying an X- or H-gate to a J-gate are displayed in Equation~\ref{eq:ssp-qidxh-properties},
\begin{equation} \label{eq:ssp-qidxh-properties}
    \begin{matrix}
        X|i\rangle = -|-i\rangle & H|i\rangle = |i\rangle \\
        X|-i\rangle = -|i\rangle  & H|-i\rangle = |-i\rangle
    \end{matrix}
\end{equation}
and the J-gate has the properties depicted in Equation~\ref{eq:ssp-qidj}.
Interestingly, the Qid-J-gate unitary ended up being equal to the negative Qid-X-gate with reversed columns and
vice versa, as seen in Equation~\ref{eq:ssp-qidx-qidj-unitary}.
\begin{equation} \label{eq:ssp-qidj}
    \begin{split}
        J|0\rangle = |i\rangle  \\
        J|1\rangle = |-i\rangle \\
        J|+\rangle = |-\rangle  \\
        J|-\rangle = |+\rangle  \\
        J|i\rangle = -|0\rangle \\
        J|-i\rangle = -|1\rangle
    \end{split}
\end{equation}

\begin{equation} \label{eq:ssp-qidx-qidj-unitary}
    QidX_2 =
    \begin{pmatrix}
        0  & 0 & 0 & 1 & 0 & 0  \\
        0  & 1 & 0 & 0 & 0 & 0  \\
        0  & 0 & 0 & 0 & 0 & -1 \\
        1  & 0 & 0 & 0 & 0 & 0  \\
        0  & 0 & 0 & 0 & -1 & 0 \\
        0  & 0 & -1 & 0 & 0 & 0
    \end{pmatrix}
    \qquad  % Insert space between matrices.
    QidJ =
    \begin{pmatrix}
        0 & 0 & -1 & 0 & 0 & 0 \\
        0 & 0 & 0 & 0 & -1 & 0 \\
        1 & 0 & 0 & 0 & 0 & 0  \\
        0 & 0 & 0 & 0 & 0 & -1 \\
        0 & 1 & 0 & 0 & 0 & 0  \\
        0 & 0 & 0 & 1 & 0 & 0
    \end{pmatrix}
\end{equation}

\begin{figure}[h!]
    \centering
    \caption{SSP conversion gate.}
    \label{fig:ssp-conv-circ}
    \scalebox{1.0}{
        \Qcircuit @C=1.0em @R=0.2em @!R { \\
        \nghost{Qudit_6 :  } & \lstick{Qudit_6 :  } & \multigate{2}{\mathrm{QidConv_3}} & \qw & \qw & \qw\\
        \nghost{Qudit_3 :  } & \lstick{Qudit_3 :  } & \ghost{\mathrm{QidConv_3}} & \ctrl{1} & \qw & \qw\\
        \nghost{Qubit :  } & \lstick{Qubit :  } & \ghost{\mathrm{QidConv_3}} & \gate{\mathrm{CHJ}} & \qw & \qw\\
        \\ }}
\end{figure}
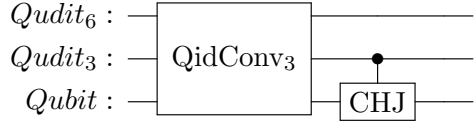

\paragraph{SSP Conversion Gate.} A gate that converts a \ndimensional{6} qudit to a \ndimensional{2} qubit when
modeling \gls{ssp} is depicted in Figure~\ref{fig:ssp-conv-circ}.
The gate uses a \ndimensional{3} qudit, or qutrit, to determine the encoding basis to apply to the qubit using a
\gls{chjgate}.
Appendix~\ref{app:conv-unitaries} and Equation~\ref{eq:ssp-chj} depict the unitary for \QidConvSSP~and the CHJ-gate
respectively.
As is demonstrated in Equation~\ref{eq:ssp-chj}, an \gls{hgate} is applied if the qutrit is in the state $|1\rangle$,
whereas a J-gate is applied if it is in the state $|2\rangle$.

\begin{equation} \label{eq:ssp-chj}
    \begin{pmatrix}
        1 & 0 & 0 & 0 & 0 & 0 \\
        0 & 1 & 0 & 0 & 0 & 0 \\
        0 & 0 & \frac{1}{\sqrt{2}} & \frac{1}{\sqrt{2}} & 0 & 0  \\
        0 & 0 & \frac{1}{\sqrt{2}} & -\frac{1}{\sqrt{2}} & 0 & 0 \\
        0 & 0 & 0 & 0 & \frac{i}{\sqrt{2}} & -\frac{1}{\sqrt{2}} \\
        0 & 0 & 0 & 0 & \frac{1}{\sqrt{2}} & -\frac{i}{\sqrt{2}} \\
    \end{pmatrix}
\end{equation}

%% file: sections/results/results.tex
\section{Results} \label{sec:results}

Tables~\ref{tab:results-bb84},~\ref{tab:results-hdbb84}, and~\ref{tab:results-ssp} list the results of each scenario
when performed 25 times.
This includes the size of the sifted key compared to the total amount of bits sent between Alice and Bob alongside the
average \gls{qber} both with and without the presence of Eve.
They also list Eve's knowledge of Alice's and Bob's sifted key.
Finally, they display the percentage of tests that produced sifted keys identical to that created in the control group
without the presence of Eve (Matches).
Results measured in Eve's presence are denoted by $E$.

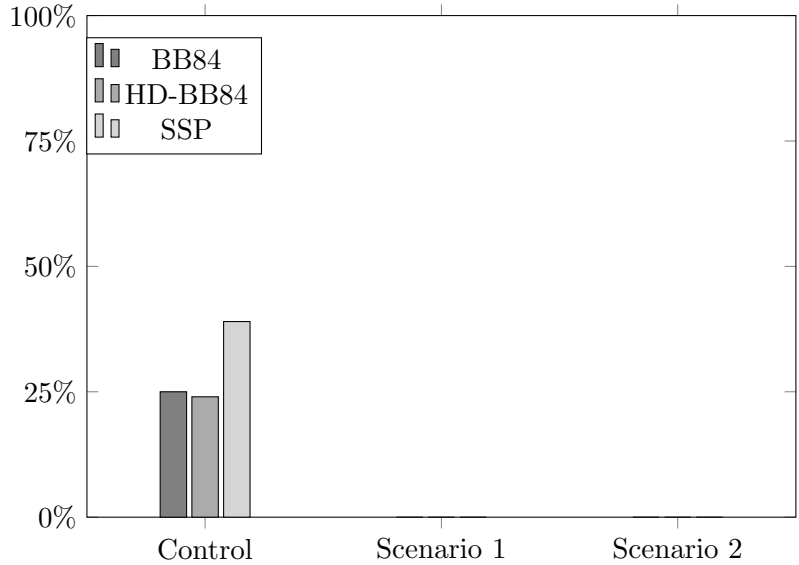
\begin{figure}[h!]
    \caption{The average QBER of BB84, its high-dimensional variant, and SSP when an eavesdropper is present.}
    \label{fig:results-qber}
    \begin{center}
        \begin{tikzpicture}
            \begin{axis} [
                ybar,
                ymin=0,
                ymax=1,
                ytick={0.0, 0.25, 0.5, 0.75, 1.0},
                yticklabels={0\%, 25\%, 50\%, 75\%, 100\%},
                symbolic x coords={Control, \scenarioANorm, \scenarioBNorm},
                xtick={Control, \scenarioANorm, \scenarioBNorm},
                enlarge x limits = 0.25,
                width=\textwidth/1.5,
                height=\textwidth/2,
                legend style={at={(-0.001, 0.84)}, anchor=west},
            ]
                \addplot [nthBarOfThree=1] table [x=test, y=BB84]{\qberData};
                \addplot [nthBarOfThree=2] table [x=test, y=HD-BB84]{\qberData};
                \addplot [nthBarOfThree=3] table [x=test, y=SSP]{\qberData};
                \legend{BB84, HD-BB84, SSP}
            \end{axis}
        \end{tikzpicture}
    \end{center}
\end{figure}

A match of 100\% in \scenarioA~and \scenarioB~without the presence of Eve indicates that the model accurately reflects
the expected behavior of the associated \gls{qkd} algorithm.
A match of 100\% in \scenarioA~and \scenarioB~with the presence of Eve indicates that the high-dimensional eavesdropping
attack produced no \gls{qber} in any test.
The average \gls{qber} produced by Eve in each scenario is illustrated in Figure~\ref{fig:results-qber}.

\begin{table}[h!]
    \caption{Results from simulating BB84.}
    \label{tab:results-bb84}
    % Length of Key + percentage of bits sent that were used as keys
    % Average QBER
    % Average Alice Knowledge
    % Average Bob Knowledge
    % Number of keys generated by Alice and Bob matched the control without Eve
    \begin{tabularx}{\linewidth}{cccccc}
                       & Key Size          & QBER    & \makecell{Alice\\Knowledge} & \makecell{Bob\\Knowledge} & Matches \\ \cline{2-6}
        Control        & \ksize{109}{54.5} & 0\%                                                                         \\
        \scenarioA     & \ksize{109}{54.5} & 0\%     &                             &                           & 100\%   \\
        \scenarioB     & \ksize{109}{54.5} & 0\%     &                             &                           & 100\%   \\
        Control$_E$    & \ksize{109}{54.5} & 25.03\% & 74.86\%                     & 75.96\%                   & 0\%     \\
        \scenarioA$_E$ & \ksize{109}{54.5} & 0\%     & 100\%                       & 100\%                     & 100\%   \\
        \scenarioB$_E$ & \ksize{109}{54.5} & 0\%     & 100\%                       & 100\%                     & 100\%   \\
    \end{tabularx}
\end{table}

\begin{table}[h!]
    \caption{Results from simulating HD-BB84.}
    \label{tab:results-hdbb84}
    \begin{tabularx}{\linewidth}{cccccc}
                       & Key Size       & QBER    & \makecell{Alice\\Knowledge} & \makecell{Bob\\Knowledge} & Matches \\ \cline{2-6}
        Control        & \ksize{78}{39} & 0\%                                                                         \\
        \scenarioA     & \ksize{78}{39} & 0\%     &                             &                           & 100\%   \\
        \scenarioB     & \ksize{78}{39} & 0\%     &                             &                           & 100\%   \\
        Control$_E$    & \ksize{78}{39} & 23.69\% & 76.26\%                     & 77.9\%                    & 0\%     \\
        \scenarioA$_E$ & \ksize{78}{39} & 0\%     & 100\%                       & 100\%                     & 100\%   \\
        \scenarioB$_E$ & \ksize{78}{39} & 0\%     & 100\%                       & 100\%                     & 100\%   \\
    \end{tabularx}
\end{table}

\begin{table}[h!]
    \caption{Results from simulating SSP.}
    \label{tab:results-ssp}
    \begin{tabularx}{\linewidth}{cccccc}
                       & Key Size         & QBER    & \makecell{Alice\\Knowledge} & \makecell{Bob\\Knowledge} & Matches \\ \cline{2-6}
        Control        & \ksize{69}{34.5} & 0\%                                                                         \\
        \scenarioA     & \ksize{69}{34.5} & 0\%     &                             &                           & 100\%   \\
        \scenarioB     & \ksize{69}{34.5} & 0\%     &                             &                           & 100\%   \\
        Control$_E$    & \ksize{69}{34.5} & 38.55\% & 71.42\%                     & 60.7\%                    & 0\%     \\
        \scenarioA$_E$ & \ksize{69}{34.5} & 0\%     & 100\%                       & 100\%                     & 100\%   \\
        \scenarioB$_E$ & \ksize{69}{34.5} & 0\%     & 100\%                       & 100\%                     & 100\%   \\
    \end{tabularx}
\end{table}

% Data for bar chart (Average QBER of Eve)
\pgfplotstableread[row sep=\\,col sep=&]{
    test       & BB84 & HD-BB84 & SSP  \\
    Control    & 0.25 & 0.24    & 0.39 \\
    \scenarioANorm & 0    & 0       & 0    \\
    \scenarioBNorm & 0    & 0       & 0    \\
}\qberData

The average \gls{qber} produced by Eve in the control group closely matches the expected \gls{qber} of the control's
corresponding algorithm.
BB84 and its high-dimensional variant should both produce a \gls{qber} of approximately 25\% while Eve is
present~\cite{Bennett1984, BechmannPasquinucci2000}, whereas \gls{ssp} should produce a \gls{qber} of approximately
33\%~\cite{BechmannPasquinucci1999}.
These results demonstrate that the presence of Eve did not introduce any \gls{qber} when using a high-dimensional
eavesdropping strategy, thus supporting the hypothesis that the use of a high-dimensional basis by Eve will produce a
minimal \gls{qber} when eavesdropping BB84 and its variants.

%% file: sections/discussion/discussion.tex
\section{Discussion} \label{sec:discussion}

Based on the results of the simulations, a potential attack vector exists for \gls{qkd} algorithms that rely on
Heisenberg's uncertainty principle to provide security.
If an eavesdropper were to measure the quantum channel in a number of dimensions equal to that in
Equation~\ref{eq:hypothesis}, no noise should be introduced by the eavesdropper.
A means of mitigating this would be the use of additional dimensions and encoding bases, thus increasing the number of
dimensions an eavesdropper would need to measure with a single sensor.
Indeed, the high-dimensional variant of BB84 would require Eve to use an \ndimensional{8} sensor, whereas \gls{ssp}
would require a \ndimensional{6} sensor.
However, this causes a reduced key efficiency, as Alice and Bob will be more likely to choose an incorrect
encoding basis~\cite{BechmannPasquinucci2000, BechmannPasquinucci1999}.

Alternatively, \gls{qkd} algorithms that rely on entanglement, such as E91, may not be affected by this attack.
E91 provides security by analyzing how entangled particles are.
This is achieved using the \gls{chsh} correlation value of the results measured by Alice and Bob.
A \gls{chsh} correlation value that differs from $-2\sqrt{2}$ indicates a lack of entanglement and the presence of an
eavesdropper~\cite{Ekert1991}.

It should also be noted that this assumes the physical possibility of a high-dimensional eavesdropping attack.
In addition to this, the tests performed could not account for natural noise that may exist in quantum channels.
A number of external sources of noise exist that can occur without the presence of Eve, including decoherence and
relaxation.
Decoherence occurs when a quantum particle spontaneously loses its state, while relaxation occurs when a particle
loses energy.
Both sources of error are more likely to occur over time.
As such, qubits are limited by their lifespan.

Another source of noise can be that caused by erroneous gate operations and measurements.
Quantum gates cause gate errors when they fail to properly alter the state of a particle, whereas a measurement error
occurs when the incorrect state is measured~\cite{Saki2021}.

While these factors could not be analyzed given time constraints and a lack of hardware, they could potentially
introduce unexpected behavior.
The qubits on the quantum channel would have a shorter lifespan when in the presence of Eve since they would have to
travel a shorter distance from Alice to Eve and from Eve to Bob.
This is because once Eve intercepts a qubit from Alice, Eve must then create a new qubit to be sent to Bob.
Therefore, the shorter lifespan of qubits should reduce the noise caused from decoherence and relaxation.
However, Eve may still produce noise depending on how accurate their measurements and gate operations are.
While this promotes the need to perform further research using physical hardware, it is unlikely that the sources of
noise provided and reduced by Eve would be significant enough to detect.

% Conclusion:
% The tests performed demonstrate the feasibility of a hypothetical attack on \QKD algorithms based on BB84
% Further research: Test on phys hardware and E91

%% file: sections/conclusion/conclusion.tex
% https://www.scribbr.com/dissertation/write-conclusion/

\section{Conclusion} \label{sec:conclusion}

A novel high-dimensional eavesdropping attack was developed and performed on \gls{qkd} algorithms that rely on
Heisenberg's uncertainty principle to provide security.
Based on simulated attacks on BB84, a high-dimensional variant of BB84, and \gls{ssp}, it was demonstrated that a
high-dimensional eavesdropping attack will produce zero \gls{qber} in a quantum channel.
For this attack to be performed, an eavesdropper must measure the quantum channel as a $d$ dimensional qudit, wherein
$d$ is determined by the dimensions of the quantum channel and number of encoding bases used.
The number of dimensions an eavesdropper must use is given in Equation~\ref{eq:hypothesis}.

This demonstrates a here-to-fore unexplored vulnerability of \gls{qkd} algorithms.
Such a vulnerability warrants research on physical hardware, as the use of a simulation could not account for
environmental noise, nor the accuracy of the sensor used by an eavesdropper.
While the results obtained show that \gls{hdqkd} offer protection by increasing the number of dimensions an
eavesdropper must measure, this also causes a reduced key efficiency as the number of potential encoding bases increase.
Finally, the findings invite further investigation of a high-dimensional eavesdropping attack on entanglement based
\gls{qkd} algorithms such as E91 as a means of protection from this attack.
\expandupon[inline]{Explore alternate means of detecting Eve other than QBER?}

%% file: appendices/result-format.tex
\begin{table}[h!]
    \caption{A table depicting the binary equivalent of a qudit measurement made by Bob and Eve when using BB84,
        wherein $\rho$ is a random bit. Bob's binary values are only used in \scenarioA.}
    \label{tab:bb84-result-format}
    \begin{tabular}{c c c}
        Measurement & Eve Binary & Bob Binary \\
        0           & 0          & 0          \\
        1           & 0          & $\rho$     \\
        2           & 1          & 1          \\
        3           & 1          & $\rho$
    \end{tabular}
\end{table}

\begin{table}[h!]
    \caption{A table depicting the binary equivalent of a qudit measurement made by Bob and Eve when using the
    high-dimensional variant of BB84, wherein $\rho$ are two random bits. Bob's binary values are only used
    in \scenarioA.}
    \label{tab:hdbb84-result-format}
    \begin{tabular}{c c c}
        Measurement & Eve Binary & Bob Binary \\
        0           & 00         & 00         \\
        1           & 00         & $\rho$     \\
        2           & 01         & 01         \\
        3           & 01         & $\rho$     \\
        4           & 10         & 10         \\
        5           & 10         & $\rho$     \\
        6           & 11         & 11         \\
        7           & 11         & $\rho$
    \end{tabular}
\end{table}

\begin{table}[h!]
    \caption{A table depicting the binary equivalent of a qudit measurement made by Bob and Eve when using SSP,
        wherein $\rho$ is a random bit. Bob's binary values are only used in \scenarioA.}
    \label{tab:ssp-result-format}
    \begin{tabular}{c c c}
        Measurement & Eve Binary & Bob Binary \\
        0           & 0          & 0          \\
        1           & 0          & $\rho$     \\
        2           & 0          & $\rho$     \\
        3           & 1          & 1          \\
        4           & 1          & $\rho$     \\
        5           & 1          & $\rho$
    \end{tabular}
\end{table}

%% file: appendices/conv-unitaries.tex
\setcounter{MaxMatrixCols}{64}

% BB84 Qid 4 Conv
The unitary matrix for \QidConvBB.
This gate is used to convert \ndimensional{4} qudits to qubits when analyzing BB84.
\begin{center}
   $\begin{pmatrix}
        1 & 0 & 0 & 0 & 0 & 0 & 0 & 0 & 0 & 0 & 0 & 0 & 0 & 0 & 0 & 0 \\
        0 & 1 & 0 & 0 & 0 & 0 & 0 & 0 & 0 & 0 & 0 & 0 & 0 & 0 & 0 & 0 \\
        0 & 0 & 1 & 0 & 0 & 0 & 0 & 0 & 0 & 0 & 0 & 0 & 0 & 0 & 0 & 0 \\
        0 & 0 & 0 & 1 & 0 & 0 & 0 & 0 & 0 & 0 & 0 & 0 & 0 & 0 & 0 & 0 \\
        0 & 0 & 0 & 0 & 0 & 1 & 0 & 0 & 0 & 0 & 0 & 0 & 0 & 0 & 0 & 0 \\
        0 & 0 & 0 & 0 & 1 & 0 & 0 & 0 & 0 & 0 & 0 & 0 & 0 & 0 & 0 & 0 \\
        0 & 0 & 0 & 0 & 0 & 0 & 0 & 1 & 0 & 0 & 0 & 0 & 0 & 0 & 0 & 0 \\
        0 & 0 & 0 & 0 & 0 & 0 & 1 & 0 & 0 & 0 & 0 & 0 & 0 & 0 & 0 & 0 \\
        0 & 0 & 0 & 0 & 0 & 0 & 0 & 0 & 0 & 0 & 1 & 0 & 0 & 0 & 0 & 0 \\
        0 & 0 & 0 & 0 & 0 & 0 & 0 & 0 & 0 & 0 & 0 & 1 & 0 & 0 & 0 & 0 \\
        0 & 0 & 0 & 0 & 0 & 0 & 0 & 0 & 1 & 0 & 0 & 0 & 0 & 0 & 0 & 0 \\
        0 & 0 & 0 & 0 & 0 & 0 & 0 & 0 & 0 & 1 & 0 & 0 & 0 & 0 & 0 & 0 \\
        0 & 0 & 0 & 0 & 0 & 0 & 0 & 0 & 0 & 0 & 0 & 0 & 0 & 0 & 0 & 1 \\
        0 & 0 & 0 & 0 & 0 & 0 & 0 & 0 & 0 & 0 & 0 & 0 & 0 & 0 & 1 & 0 \\
        0 & 0 & 0 & 0 & 0 & 0 & 0 & 0 & 0 & 0 & 0 & 0 & 0 & 1 & 0 & 0 \\
        0 & 0 & 0 & 0 & 0 & 0 & 0 & 0 & 0 & 0 & 0 & 0 & 1 & 0 & 0 & 0
   \end{pmatrix}$
\end{center}

% HD-BB84 Qid 8 Conv
The unitary matrix for \QidConvHDBB~is depicted below.
This gate is used to convert \ndimensional{8} qudits to \ndimensional{4} qudits when analyzing the high-dimensional
variant of BB84.
\begin{center}
   \makebox[\textwidth]{%
       \resizebox{0.99\paperwidth}{!}{%
           $\begin{pmatrix}
                1 & 0 & 0 & 0 & 0 & 0 & 0 & 0 & 0 & 0 & 0 & 0 & 0 & 0 & 0 & 0 & 0 & 0 & 0 & 0 & 0 & 0 & 0 & 0 & 0 & 0 & 0 & 0 & 0 & 0 & 0 & 0 & 0 & 0 & 0 & 0 & 0 & 0 & 0 & 0 & 0 & 0 & 0 & 0 & 0 & 0 & 0 & 0 & 0 & 0 & 0 & 0 & 0 & 0 & 0 & 0 & 0 & 0 & 0 & 0 & 0 & 0 & 0 & 0 \\
                0 & 1 & 0 & 0 & 0 & 0 & 0 & 0 & 0 & 0 & 0 & 0 & 0 & 0 & 0 & 0 & 0 & 0 & 0 & 0 & 0 & 0 & 0 & 0 & 0 & 0 & 0 & 0 & 0 & 0 & 0 & 0 & 0 & 0 & 0 & 0 & 0 & 0 & 0 & 0 & 0 & 0 & 0 & 0 & 0 & 0 & 0 & 0 & 0 & 0 & 0 & 0 & 0 & 0 & 0 & 0 & 0 & 0 & 0 & 0 & 0 & 0 & 0 & 0 \\
                0 & 0 & 1 & 0 & 0 & 0 & 0 & 0 & 0 & 0 & 0 & 0 & 0 & 0 & 0 & 0 & 0 & 0 & 0 & 0 & 0 & 0 & 0 & 0 & 0 & 0 & 0 & 0 & 0 & 0 & 0 & 0 & 0 & 0 & 0 & 0 & 0 & 0 & 0 & 0 & 0 & 0 & 0 & 0 & 0 & 0 & 0 & 0 & 0 & 0 & 0 & 0 & 0 & 0 & 0 & 0 & 0 & 0 & 0 & 0 & 0 & 0 & 0 & 0 \\
                0 & 0 & 0 & 1 & 0 & 0 & 0 & 0 & 0 & 0 & 0 & 0 & 0 & 0 & 0 & 0 & 0 & 0 & 0 & 0 & 0 & 0 & 0 & 0 & 0 & 0 & 0 & 0 & 0 & 0 & 0 & 0 & 0 & 0 & 0 & 0 & 0 & 0 & 0 & 0 & 0 & 0 & 0 & 0 & 0 & 0 & 0 & 0 & 0 & 0 & 0 & 0 & 0 & 0 & 0 & 0 & 0 & 0 & 0 & 0 & 0 & 0 & 0 & 0 \\
                0 & 0 & 0 & 0 & 1 & 0 & 0 & 0 & 0 & 0 & 0 & 0 & 0 & 0 & 0 & 0 & 0 & 0 & 0 & 0 & 0 & 0 & 0 & 0 & 0 & 0 & 0 & 0 & 0 & 0 & 0 & 0 & 0 & 0 & 0 & 0 & 0 & 0 & 0 & 0 & 0 & 0 & 0 & 0 & 0 & 0 & 0 & 0 & 0 & 0 & 0 & 0 & 0 & 0 & 0 & 0 & 0 & 0 & 0 & 0 & 0 & 0 & 0 & 0 \\
                0 & 0 & 0 & 0 & 0 & 1 & 0 & 0 & 0 & 0 & 0 & 0 & 0 & 0 & 0 & 0 & 0 & 0 & 0 & 0 & 0 & 0 & 0 & 0 & 0 & 0 & 0 & 0 & 0 & 0 & 0 & 0 & 0 & 0 & 0 & 0 & 0 & 0 & 0 & 0 & 0 & 0 & 0 & 0 & 0 & 0 & 0 & 0 & 0 & 0 & 0 & 0 & 0 & 0 & 0 & 0 & 0 & 0 & 0 & 0 & 0 & 0 & 0 & 0 \\
                0 & 0 & 0 & 0 & 0 & 0 & 1 & 0 & 0 & 0 & 0 & 0 & 0 & 0 & 0 & 0 & 0 & 0 & 0 & 0 & 0 & 0 & 0 & 0 & 0 & 0 & 0 & 0 & 0 & 0 & 0 & 0 & 0 & 0 & 0 & 0 & 0 & 0 & 0 & 0 & 0 & 0 & 0 & 0 & 0 & 0 & 0 & 0 & 0 & 0 & 0 & 0 & 0 & 0 & 0 & 0 & 0 & 0 & 0 & 0 & 0 & 0 & 0 & 0 \\
                0 & 0 & 0 & 0 & 0 & 0 & 0 & 1 & 0 & 0 & 0 & 0 & 0 & 0 & 0 & 0 & 0 & 0 & 0 & 0 & 0 & 0 & 0 & 0 & 0 & 0 & 0 & 0 & 0 & 0 & 0 & 0 & 0 & 0 & 0 & 0 & 0 & 0 & 0 & 0 & 0 & 0 & 0 & 0 & 0 & 0 & 0 & 0 & 0 & 0 & 0 & 0 & 0 & 0 & 0 & 0 & 0 & 0 & 0 & 0 & 0 & 0 & 0 & 0 \\
                0 & 0 & 0 & 0 & 0 & 0 & 0 & 0 & 0 & 0 & 0 & 0 & 1 & 0 & 0 & 0 & 0 & 0 & 0 & 0 & 0 & 0 & 0 & 0 & 0 & 0 & 0 & 0 & 0 & 0 & 0 & 0 & 0 & 0 & 0 & 0 & 0 & 0 & 0 & 0 & 0 & 0 & 0 & 0 & 0 & 0 & 0 & 0 & 0 & 0 & 0 & 0 & 0 & 0 & 0 & 0 & 0 & 0 & 0 & 0 & 0 & 0 & 0 & 0 \\
                0 & 0 & 0 & 0 & 0 & 0 & 0 & 0 & 0 & 0 & 0 & 0 & 0 & 1 & 0 & 0 & 0 & 0 & 0 & 0 & 0 & 0 & 0 & 0 & 0 & 0 & 0 & 0 & 0 & 0 & 0 & 0 & 0 & 0 & 0 & 0 & 0 & 0 & 0 & 0 & 0 & 0 & 0 & 0 & 0 & 0 & 0 & 0 & 0 & 0 & 0 & 0 & 0 & 0 & 0 & 0 & 0 & 0 & 0 & 0 & 0 & 0 & 0 & 0 \\
                0 & 0 & 0 & 0 & 0 & 0 & 0 & 0 & 0 & 0 & 0 & 0 & 0 & 0 & 1 & 0 & 0 & 0 & 0 & 0 & 0 & 0 & 0 & 0 & 0 & 0 & 0 & 0 & 0 & 0 & 0 & 0 & 0 & 0 & 0 & 0 & 0 & 0 & 0 & 0 & 0 & 0 & 0 & 0 & 0 & 0 & 0 & 0 & 0 & 0 & 0 & 0 & 0 & 0 & 0 & 0 & 0 & 0 & 0 & 0 & 0 & 0 & 0 & 0 \\
                0 & 0 & 0 & 0 & 0 & 0 & 0 & 0 & 0 & 0 & 0 & 0 & 0 & 0 & 0 & 1 & 0 & 0 & 0 & 0 & 0 & 0 & 0 & 0 & 0 & 0 & 0 & 0 & 0 & 0 & 0 & 0 & 0 & 0 & 0 & 0 & 0 & 0 & 0 & 0 & 0 & 0 & 0 & 0 & 0 & 0 & 0 & 0 & 0 & 0 & 0 & 0 & 0 & 0 & 0 & 0 & 0 & 0 & 0 & 0 & 0 & 0 & 0 & 0 \\
                0 & 0 & 0 & 0 & 0 & 0 & 0 & 0 & 1 & 0 & 0 & 0 & 0 & 0 & 0 & 0 & 0 & 0 & 0 & 0 & 0 & 0 & 0 & 0 & 0 & 0 & 0 & 0 & 0 & 0 & 0 & 0 & 0 & 0 & 0 & 0 & 0 & 0 & 0 & 0 & 0 & 0 & 0 & 0 & 0 & 0 & 0 & 0 & 0 & 0 & 0 & 0 & 0 & 0 & 0 & 0 & 0 & 0 & 0 & 0 & 0 & 0 & 0 & 0 \\
                0 & 0 & 0 & 0 & 0 & 0 & 0 & 0 & 0 & 1 & 0 & 0 & 0 & 0 & 0 & 0 & 0 & 0 & 0 & 0 & 0 & 0 & 0 & 0 & 0 & 0 & 0 & 0 & 0 & 0 & 0 & 0 & 0 & 0 & 0 & 0 & 0 & 0 & 0 & 0 & 0 & 0 & 0 & 0 & 0 & 0 & 0 & 0 & 0 & 0 & 0 & 0 & 0 & 0 & 0 & 0 & 0 & 0 & 0 & 0 & 0 & 0 & 0 & 0 \\
                0 & 0 & 0 & 0 & 0 & 0 & 0 & 0 & 0 & 0 & 1 & 0 & 0 & 0 & 0 & 0 & 0 & 0 & 0 & 0 & 0 & 0 & 0 & 0 & 0 & 0 & 0 & 0 & 0 & 0 & 0 & 0 & 0 & 0 & 0 & 0 & 0 & 0 & 0 & 0 & 0 & 0 & 0 & 0 & 0 & 0 & 0 & 0 & 0 & 0 & 0 & 0 & 0 & 0 & 0 & 0 & 0 & 0 & 0 & 0 & 0 & 0 & 0 & 0 \\
                0 & 0 & 0 & 0 & 0 & 0 & 0 & 0 & 0 & 0 & 0 & 1 & 0 & 0 & 0 & 0 & 0 & 0 & 0 & 0 & 0 & 0 & 0 & 0 & 0 & 0 & 0 & 0 & 0 & 0 & 0 & 0 & 0 & 0 & 0 & 0 & 0 & 0 & 0 & 0 & 0 & 0 & 0 & 0 & 0 & 0 & 0 & 0 & 0 & 0 & 0 & 0 & 0 & 0 & 0 & 0 & 0 & 0 & 0 & 0 & 0 & 0 & 0 & 0 \\
                0 & 0 & 0 & 0 & 0 & 0 & 0 & 0 & 0 & 0 & 0 & 0 & 0 & 0 & 0 & 0 & 0 & 1 & 0 & 0 & 0 & 0 & 0 & 0 & 0 & 0 & 0 & 0 & 0 & 0 & 0 & 0 & 0 & 0 & 0 & 0 & 0 & 0 & 0 & 0 & 0 & 0 & 0 & 0 & 0 & 0 & 0 & 0 & 0 & 0 & 0 & 0 & 0 & 0 & 0 & 0 & 0 & 0 & 0 & 0 & 0 & 0 & 0 & 0 \\
                0 & 0 & 0 & 0 & 0 & 0 & 0 & 0 & 0 & 0 & 0 & 0 & 0 & 0 & 0 & 0 & 1 & 0 & 0 & 0 & 0 & 0 & 0 & 0 & 0 & 0 & 0 & 0 & 0 & 0 & 0 & 0 & 0 & 0 & 0 & 0 & 0 & 0 & 0 & 0 & 0 & 0 & 0 & 0 & 0 & 0 & 0 & 0 & 0 & 0 & 0 & 0 & 0 & 0 & 0 & 0 & 0 & 0 & 0 & 0 & 0 & 0 & 0 & 0 \\
                0 & 0 & 0 & 0 & 0 & 0 & 0 & 0 & 0 & 0 & 0 & 0 & 0 & 0 & 0 & 0 & 0 & 0 & 0 & 1 & 0 & 0 & 0 & 0 & 0 & 0 & 0 & 0 & 0 & 0 & 0 & 0 & 0 & 0 & 0 & 0 & 0 & 0 & 0 & 0 & 0 & 0 & 0 & 0 & 0 & 0 & 0 & 0 & 0 & 0 & 0 & 0 & 0 & 0 & 0 & 0 & 0 & 0 & 0 & 0 & 0 & 0 & 0 & 0 \\
                0 & 0 & 0 & 0 & 0 & 0 & 0 & 0 & 0 & 0 & 0 & 0 & 0 & 0 & 0 & 0 & 0 & 0 & 1 & 0 & 0 & 0 & 0 & 0 & 0 & 0 & 0 & 0 & 0 & 0 & 0 & 0 & 0 & 0 & 0 & 0 & 0 & 0 & 0 & 0 & 0 & 0 & 0 & 0 & 0 & 0 & 0 & 0 & 0 & 0 & 0 & 0 & 0 & 0 & 0 & 0 & 0 & 0 & 0 & 0 & 0 & 0 & 0 & 0 \\
                0 & 0 & 0 & 0 & 0 & 0 & 0 & 0 & 0 & 0 & 0 & 0 & 0 & 0 & 0 & 0 & 0 & 0 & 0 & 0 & 0 & 1 & 0 & 0 & 0 & 0 & 0 & 0 & 0 & 0 & 0 & 0 & 0 & 0 & 0 & 0 & 0 & 0 & 0 & 0 & 0 & 0 & 0 & 0 & 0 & 0 & 0 & 0 & 0 & 0 & 0 & 0 & 0 & 0 & 0 & 0 & 0 & 0 & 0 & 0 & 0 & 0 & 0 & 0 \\
                0 & 0 & 0 & 0 & 0 & 0 & 0 & 0 & 0 & 0 & 0 & 0 & 0 & 0 & 0 & 0 & 0 & 0 & 0 & 0 & 1 & 0 & 0 & 0 & 0 & 0 & 0 & 0 & 0 & 0 & 0 & 0 & 0 & 0 & 0 & 0 & 0 & 0 & 0 & 0 & 0 & 0 & 0 & 0 & 0 & 0 & 0 & 0 & 0 & 0 & 0 & 0 & 0 & 0 & 0 & 0 & 0 & 0 & 0 & 0 & 0 & 0 & 0 & 0 \\
                0 & 0 & 0 & 0 & 0 & 0 & 0 & 0 & 0 & 0 & 0 & 0 & 0 & 0 & 0 & 0 & 0 & 0 & 0 & 0 & 0 & 0 & 0 & 1 & 0 & 0 & 0 & 0 & 0 & 0 & 0 & 0 & 0 & 0 & 0 & 0 & 0 & 0 & 0 & 0 & 0 & 0 & 0 & 0 & 0 & 0 & 0 & 0 & 0 & 0 & 0 & 0 & 0 & 0 & 0 & 0 & 0 & 0 & 0 & 0 & 0 & 0 & 0 & 0 \\
                0 & 0 & 0 & 0 & 0 & 0 & 0 & 0 & 0 & 0 & 0 & 0 & 0 & 0 & 0 & 0 & 0 & 0 & 0 & 0 & 0 & 0 & 1 & 0 & 0 & 0 & 0 & 0 & 0 & 0 & 0 & 0 & 0 & 0 & 0 & 0 & 0 & 0 & 0 & 0 & 0 & 0 & 0 & 0 & 0 & 0 & 0 & 0 & 0 & 0 & 0 & 0 & 0 & 0 & 0 & 0 & 0 & 0 & 0 & 0 & 0 & 0 & 0 & 0 \\
                0 & 0 & 0 & 0 & 0 & 0 & 0 & 0 & 0 & 0 & 0 & 0 & 0 & 0 & 0 & 0 & 0 & 0 & 0 & 0 & 0 & 0 & 0 & 0 & 0 & 0 & 0 & 0 & 0 & 1 & 0 & 0 & 0 & 0 & 0 & 0 & 0 & 0 & 0 & 0 & 0 & 0 & 0 & 0 & 0 & 0 & 0 & 0 & 0 & 0 & 0 & 0 & 0 & 0 & 0 & 0 & 0 & 0 & 0 & 0 & 0 & 0 & 0 & 0 \\
                0 & 0 & 0 & 0 & 0 & 0 & 0 & 0 & 0 & 0 & 0 & 0 & 0 & 0 & 0 & 0 & 0 & 0 & 0 & 0 & 0 & 0 & 0 & 0 & 0 & 0 & 0 & 0 & 1 & 0 & 0 & 0 & 0 & 0 & 0 & 0 & 0 & 0 & 0 & 0 & 0 & 0 & 0 & 0 & 0 & 0 & 0 & 0 & 0 & 0 & 0 & 0 & 0 & 0 & 0 & 0 & 0 & 0 & 0 & 0 & 0 & 0 & 0 & 0 \\
                0 & 0 & 0 & 0 & 0 & 0 & 0 & 0 & 0 & 0 & 0 & 0 & 0 & 0 & 0 & 0 & 0 & 0 & 0 & 0 & 0 & 0 & 0 & 0 & 0 & 0 & 0 & 0 & 0 & 0 & 0 & 1 & 0 & 0 & 0 & 0 & 0 & 0 & 0 & 0 & 0 & 0 & 0 & 0 & 0 & 0 & 0 & 0 & 0 & 0 & 0 & 0 & 0 & 0 & 0 & 0 & 0 & 0 & 0 & 0 & 0 & 0 & 0 & 0 \\
                0 & 0 & 0 & 0 & 0 & 0 & 0 & 0 & 0 & 0 & 0 & 0 & 0 & 0 & 0 & 0 & 0 & 0 & 0 & 0 & 0 & 0 & 0 & 0 & 0 & 0 & 0 & 0 & 0 & 0 & 1 & 0 & 0 & 0 & 0 & 0 & 0 & 0 & 0 & 0 & 0 & 0 & 0 & 0 & 0 & 0 & 0 & 0 & 0 & 0 & 0 & 0 & 0 & 0 & 0 & 0 & 0 & 0 & 0 & 0 & 0 & 0 & 0 & 0 \\
                0 & 0 & 0 & 0 & 0 & 0 & 0 & 0 & 0 & 0 & 0 & 0 & 0 & 0 & 0 & 0 & 0 & 0 & 0 & 0 & 0 & 0 & 0 & 0 & 0 & 1 & 0 & 0 & 0 & 0 & 0 & 0 & 0 & 0 & 0 & 0 & 0 & 0 & 0 & 0 & 0 & 0 & 0 & 0 & 0 & 0 & 0 & 0 & 0 & 0 & 0 & 0 & 0 & 0 & 0 & 0 & 0 & 0 & 0 & 0 & 0 & 0 & 0 & 0 \\
                0 & 0 & 0 & 0 & 0 & 0 & 0 & 0 & 0 & 0 & 0 & 0 & 0 & 0 & 0 & 0 & 0 & 0 & 0 & 0 & 0 & 0 & 0 & 0 & 1 & 0 & 0 & 0 & 0 & 0 & 0 & 0 & 0 & 0 & 0 & 0 & 0 & 0 & 0 & 0 & 0 & 0 & 0 & 0 & 0 & 0 & 0 & 0 & 0 & 0 & 0 & 0 & 0 & 0 & 0 & 0 & 0 & 0 & 0 & 0 & 0 & 0 & 0 & 0 \\
                0 & 0 & 0 & 0 & 0 & 0 & 0 & 0 & 0 & 0 & 0 & 0 & 0 & 0 & 0 & 0 & 0 & 0 & 0 & 0 & 0 & 0 & 0 & 0 & 0 & 0 & 0 & 1 & 0 & 0 & 0 & 0 & 0 & 0 & 0 & 0 & 0 & 0 & 0 & 0 & 0 & 0 & 0 & 0 & 0 & 0 & 0 & 0 & 0 & 0 & 0 & 0 & 0 & 0 & 0 & 0 & 0 & 0 & 0 & 0 & 0 & 0 & 0 & 0 \\
                0 & 0 & 0 & 0 & 0 & 0 & 0 & 0 & 0 & 0 & 0 & 0 & 0 & 0 & 0 & 0 & 0 & 0 & 0 & 0 & 0 & 0 & 0 & 0 & 0 & 0 & 1 & 0 & 0 & 0 & 0 & 0 & 0 & 0 & 0 & 0 & 0 & 0 & 0 & 0 & 0 & 0 & 0 & 0 & 0 & 0 & 0 & 0 & 0 & 0 & 0 & 0 & 0 & 0 & 0 & 0 & 0 & 0 & 0 & 0 & 0 & 0 & 0 & 0 \\
                0 & 0 & 0 & 0 & 0 & 0 & 0 & 0 & 0 & 0 & 0 & 0 & 0 & 0 & 0 & 0 & 0 & 0 & 0 & 0 & 0 & 0 & 0 & 0 & 0 & 0 & 0 & 0 & 0 & 0 & 0 & 0 & 0 & 0 & 1 & 0 & 0 & 0 & 0 & 0 & 0 & 0 & 0 & 0 & 0 & 0 & 0 & 0 & 0 & 0 & 0 & 0 & 0 & 0 & 0 & 0 & 0 & 0 & 0 & 0 & 0 & 0 & 0 & 0 \\
                0 & 0 & 0 & 0 & 0 & 0 & 0 & 0 & 0 & 0 & 0 & 0 & 0 & 0 & 0 & 0 & 0 & 0 & 0 & 0 & 0 & 0 & 0 & 0 & 0 & 0 & 0 & 0 & 0 & 0 & 0 & 0 & 0 & 1 & 0 & 0 & 0 & 0 & 0 & 0 & 0 & 0 & 0 & 0 & 0 & 0 & 0 & 0 & 0 & 0 & 0 & 0 & 0 & 0 & 0 & 0 & 0 & 0 & 0 & 0 & 0 & 0 & 0 & 0 \\
                0 & 0 & 0 & 0 & 0 & 0 & 0 & 0 & 0 & 0 & 0 & 0 & 0 & 0 & 0 & 0 & 0 & 0 & 0 & 0 & 0 & 0 & 0 & 0 & 0 & 0 & 0 & 0 & 0 & 0 & 0 & 0 & 1 & 0 & 0 & 0 & 0 & 0 & 0 & 0 & 0 & 0 & 0 & 0 & 0 & 0 & 0 & 0 & 0 & 0 & 0 & 0 & 0 & 0 & 0 & 0 & 0 & 0 & 0 & 0 & 0 & 0 & 0 & 0 \\
                0 & 0 & 0 & 0 & 0 & 0 & 0 & 0 & 0 & 0 & 0 & 0 & 0 & 0 & 0 & 0 & 0 & 0 & 0 & 0 & 0 & 0 & 0 & 0 & 0 & 0 & 0 & 0 & 0 & 0 & 0 & 0 & 0 & 0 & 0 & 1 & 0 & 0 & 0 & 0 & 0 & 0 & 0 & 0 & 0 & 0 & 0 & 0 & 0 & 0 & 0 & 0 & 0 & 0 & 0 & 0 & 0 & 0 & 0 & 0 & 0 & 0 & 0 & 0 \\
                0 & 0 & 0 & 0 & 0 & 0 & 0 & 0 & 0 & 0 & 0 & 0 & 0 & 0 & 0 & 0 & 0 & 0 & 0 & 0 & 0 & 0 & 0 & 0 & 0 & 0 & 0 & 0 & 0 & 0 & 0 & 0 & 0 & 0 & 0 & 0 & 0 & 0 & 1 & 0 & 0 & 0 & 0 & 0 & 0 & 0 & 0 & 0 & 0 & 0 & 0 & 0 & 0 & 0 & 0 & 0 & 0 & 0 & 0 & 0 & 0 & 0 & 0 & 0 \\
                0 & 0 & 0 & 0 & 0 & 0 & 0 & 0 & 0 & 0 & 0 & 0 & 0 & 0 & 0 & 0 & 0 & 0 & 0 & 0 & 0 & 0 & 0 & 0 & 0 & 0 & 0 & 0 & 0 & 0 & 0 & 0 & 0 & 0 & 0 & 0 & 0 & 1 & 0 & 0 & 0 & 0 & 0 & 0 & 0 & 0 & 0 & 0 & 0 & 0 & 0 & 0 & 0 & 0 & 0 & 0 & 0 & 0 & 0 & 0 & 0 & 0 & 0 & 0 \\
                0 & 0 & 0 & 0 & 0 & 0 & 0 & 0 & 0 & 0 & 0 & 0 & 0 & 0 & 0 & 0 & 0 & 0 & 0 & 0 & 0 & 0 & 0 & 0 & 0 & 0 & 0 & 0 & 0 & 0 & 0 & 0 & 0 & 0 & 0 & 0 & 1 & 0 & 0 & 0 & 0 & 0 & 0 & 0 & 0 & 0 & 0 & 0 & 0 & 0 & 0 & 0 & 0 & 0 & 0 & 0 & 0 & 0 & 0 & 0 & 0 & 0 & 0 & 0 \\
                0 & 0 & 0 & 0 & 0 & 0 & 0 & 0 & 0 & 0 & 0 & 0 & 0 & 0 & 0 & 0 & 0 & 0 & 0 & 0 & 0 & 0 & 0 & 0 & 0 & 0 & 0 & 0 & 0 & 0 & 0 & 0 & 0 & 0 & 0 & 0 & 0 & 0 & 0 & 1 & 0 & 0 & 0 & 0 & 0 & 0 & 0 & 0 & 0 & 0 & 0 & 0 & 0 & 0 & 0 & 0 & 0 & 0 & 0 & 0 & 0 & 0 & 0 & 0 \\
                0 & 0 & 0 & 0 & 0 & 0 & 0 & 0 & 0 & 0 & 0 & 0 & 0 & 0 & 0 & 0 & 0 & 0 & 0 & 0 & 0 & 0 & 0 & 0 & 0 & 0 & 0 & 0 & 0 & 0 & 0 & 0 & 0 & 0 & 0 & 0 & 0 & 0 & 0 & 0 & 0 & 0 & 0 & 0 & 0 & 0 & 1 & 0 & 0 & 0 & 0 & 0 & 0 & 0 & 0 & 0 & 0 & 0 & 0 & 0 & 0 & 0 & 0 & 0 \\
                0 & 0 & 0 & 0 & 0 & 0 & 0 & 0 & 0 & 0 & 0 & 0 & 0 & 0 & 0 & 0 & 0 & 0 & 0 & 0 & 0 & 0 & 0 & 0 & 0 & 0 & 0 & 0 & 0 & 0 & 0 & 0 & 0 & 0 & 0 & 0 & 0 & 0 & 0 & 0 & 0 & 0 & 0 & 0 & 0 & 1 & 0 & 0 & 0 & 0 & 0 & 0 & 0 & 0 & 0 & 0 & 0 & 0 & 0 & 0 & 0 & 0 & 0 & 0 \\
                0 & 0 & 0 & 0 & 0 & 0 & 0 & 0 & 0 & 0 & 0 & 0 & 0 & 0 & 0 & 0 & 0 & 0 & 0 & 0 & 0 & 0 & 0 & 0 & 0 & 0 & 0 & 0 & 0 & 0 & 0 & 0 & 0 & 0 & 0 & 0 & 0 & 0 & 0 & 0 & 0 & 0 & 0 & 0 & 1 & 0 & 0 & 0 & 0 & 0 & 0 & 0 & 0 & 0 & 0 & 0 & 0 & 0 & 0 & 0 & 0 & 0 & 0 & 0 \\
                0 & 0 & 0 & 0 & 0 & 0 & 0 & 0 & 0 & 0 & 0 & 0 & 0 & 0 & 0 & 0 & 0 & 0 & 0 & 0 & 0 & 0 & 0 & 0 & 0 & 0 & 0 & 0 & 0 & 0 & 0 & 0 & 0 & 0 & 0 & 0 & 0 & 0 & 0 & 0 & 0 & 0 & 0 & 0 & 0 & 0 & 0 & 1 & 0 & 0 & 0 & 0 & 0 & 0 & 0 & 0 & 0 & 0 & 0 & 0 & 0 & 0 & 0 & 0 \\
                0 & 0 & 0 & 0 & 0 & 0 & 0 & 0 & 0 & 0 & 0 & 0 & 0 & 0 & 0 & 0 & 0 & 0 & 0 & 0 & 0 & 0 & 0 & 0 & 0 & 0 & 0 & 0 & 0 & 0 & 0 & 0 & 0 & 0 & 0 & 0 & 0 & 0 & 0 & 0 & 0 & 0 & 1 & 0 & 0 & 0 & 0 & 0 & 0 & 0 & 0 & 0 & 0 & 0 & 0 & 0 & 0 & 0 & 0 & 0 & 0 & 0 & 0 & 0 \\
                0 & 0 & 0 & 0 & 0 & 0 & 0 & 0 & 0 & 0 & 0 & 0 & 0 & 0 & 0 & 0 & 0 & 0 & 0 & 0 & 0 & 0 & 0 & 0 & 0 & 0 & 0 & 0 & 0 & 0 & 0 & 0 & 0 & 0 & 0 & 0 & 0 & 0 & 0 & 0 & 0 & 1 & 0 & 0 & 0 & 0 & 0 & 0 & 0 & 0 & 0 & 0 & 0 & 0 & 0 & 0 & 0 & 0 & 0 & 0 & 0 & 0 & 0 & 0 \\
                0 & 0 & 0 & 0 & 0 & 0 & 0 & 0 & 0 & 0 & 0 & 0 & 0 & 0 & 0 & 0 & 0 & 0 & 0 & 0 & 0 & 0 & 0 & 0 & 0 & 0 & 0 & 0 & 0 & 0 & 0 & 0 & 0 & 0 & 0 & 0 & 0 & 0 & 0 & 0 & 1 & 0 & 0 & 0 & 0 & 0 & 0 & 0 & 0 & 0 & 0 & 0 & 0 & 0 & 0 & 0 & 0 & 0 & 0 & 0 & 0 & 0 & 0 & 0 \\
                0 & 0 & 0 & 0 & 0 & 0 & 0 & 0 & 0 & 0 & 0 & 0 & 0 & 0 & 0 & 0 & 0 & 0 & 0 & 0 & 0 & 0 & 0 & 0 & 0 & 0 & 0 & 0 & 0 & 0 & 0 & 0 & 0 & 0 & 0 & 0 & 0 & 0 & 0 & 0 & 0 & 0 & 0 & 1 & 0 & 0 & 0 & 0 & 0 & 0 & 0 & 0 & 0 & 0 & 0 & 0 & 0 & 0 & 0 & 0 & 0 & 0 & 0 & 0 \\
                0 & 0 & 0 & 0 & 0 & 0 & 0 & 0 & 0 & 0 & 0 & 0 & 0 & 0 & 0 & 0 & 0 & 0 & 0 & 0 & 0 & 0 & 0 & 0 & 0 & 0 & 0 & 0 & 0 & 0 & 0 & 0 & 0 & 0 & 0 & 0 & 0 & 0 & 0 & 0 & 0 & 0 & 0 & 0 & 0 & 0 & 0 & 0 & 0 & 0 & 0 & 1 & 0 & 0 & 0 & 0 & 0 & 0 & 0 & 0 & 0 & 0 & 0 & 0 \\
                0 & 0 & 0 & 0 & 0 & 0 & 0 & 0 & 0 & 0 & 0 & 0 & 0 & 0 & 0 & 0 & 0 & 0 & 0 & 0 & 0 & 0 & 0 & 0 & 0 & 0 & 0 & 0 & 0 & 0 & 0 & 0 & 0 & 0 & 0 & 0 & 0 & 0 & 0 & 0 & 0 & 0 & 0 & 0 & 0 & 0 & 0 & 0 & 0 & 0 & 1 & 0 & 0 & 0 & 0 & 0 & 0 & 0 & 0 & 0 & 0 & 0 & 0 & 0 \\
                0 & 0 & 0 & 0 & 0 & 0 & 0 & 0 & 0 & 0 & 0 & 0 & 0 & 0 & 0 & 0 & 0 & 0 & 0 & 0 & 0 & 0 & 0 & 0 & 0 & 0 & 0 & 0 & 0 & 0 & 0 & 0 & 0 & 0 & 0 & 0 & 0 & 0 & 0 & 0 & 0 & 0 & 0 & 0 & 0 & 0 & 0 & 0 & 0 & 1 & 0 & 0 & 0 & 0 & 0 & 0 & 0 & 0 & 0 & 0 & 0 & 0 & 0 & 0 \\
                0 & 0 & 0 & 0 & 0 & 0 & 0 & 0 & 0 & 0 & 0 & 0 & 0 & 0 & 0 & 0 & 0 & 0 & 0 & 0 & 0 & 0 & 0 & 0 & 0 & 0 & 0 & 0 & 0 & 0 & 0 & 0 & 0 & 0 & 0 & 0 & 0 & 0 & 0 & 0 & 0 & 0 & 0 & 0 & 0 & 0 & 0 & 0 & 1 & 0 & 0 & 0 & 0 & 0 & 0 & 0 & 0 & 0 & 0 & 0 & 0 & 0 & 0 & 0 \\
                0 & 0 & 0 & 0 & 0 & 0 & 0 & 0 & 0 & 0 & 0 & 0 & 0 & 0 & 0 & 0 & 0 & 0 & 0 & 0 & 0 & 0 & 0 & 0 & 0 & 0 & 0 & 0 & 0 & 0 & 0 & 0 & 0 & 0 & 0 & 0 & 0 & 0 & 0 & 0 & 0 & 0 & 0 & 0 & 0 & 0 & 0 & 0 & 0 & 0 & 0 & 0 & 0 & 0 & 0 & 1 & 0 & 0 & 0 & 0 & 0 & 0 & 0 & 0 \\
                0 & 0 & 0 & 0 & 0 & 0 & 0 & 0 & 0 & 0 & 0 & 0 & 0 & 0 & 0 & 0 & 0 & 0 & 0 & 0 & 0 & 0 & 0 & 0 & 0 & 0 & 0 & 0 & 0 & 0 & 0 & 0 & 0 & 0 & 0 & 0 & 0 & 0 & 0 & 0 & 0 & 0 & 0 & 0 & 0 & 0 & 0 & 0 & 0 & 0 & 0 & 0 & 0 & 0 & 1 & 0 & 0 & 0 & 0 & 0 & 0 & 0 & 0 & 0 \\
                0 & 0 & 0 & 0 & 0 & 0 & 0 & 0 & 0 & 0 & 0 & 0 & 0 & 0 & 0 & 0 & 0 & 0 & 0 & 0 & 0 & 0 & 0 & 0 & 0 & 0 & 0 & 0 & 0 & 0 & 0 & 0 & 0 & 0 & 0 & 0 & 0 & 0 & 0 & 0 & 0 & 0 & 0 & 0 & 0 & 0 & 0 & 0 & 0 & 0 & 0 & 0 & 0 & 1 & 0 & 0 & 0 & 0 & 0 & 0 & 0 & 0 & 0 & 0 \\
                0 & 0 & 0 & 0 & 0 & 0 & 0 & 0 & 0 & 0 & 0 & 0 & 0 & 0 & 0 & 0 & 0 & 0 & 0 & 0 & 0 & 0 & 0 & 0 & 0 & 0 & 0 & 0 & 0 & 0 & 0 & 0 & 0 & 0 & 0 & 0 & 0 & 0 & 0 & 0 & 0 & 0 & 0 & 0 & 0 & 0 & 0 & 0 & 0 & 0 & 0 & 0 & 1 & 0 & 0 & 0 & 0 & 0 & 0 & 0 & 0 & 0 & 0 & 0 \\
                0 & 0 & 0 & 0 & 0 & 0 & 0 & 0 & 0 & 0 & 0 & 0 & 0 & 0 & 0 & 0 & 0 & 0 & 0 & 0 & 0 & 0 & 0 & 0 & 0 & 0 & 0 & 0 & 0 & 0 & 0 & 0 & 0 & 0 & 0 & 0 & 0 & 0 & 0 & 0 & 0 & 0 & 0 & 0 & 0 & 0 & 0 & 0 & 0 & 0 & 0 & 0 & 0 & 0 & 0 & 0 & 0 & 0 & 0 & 0 & 0 & 0 & 0 & 1 \\
                0 & 0 & 0 & 0 & 0 & 0 & 0 & 0 & 0 & 0 & 0 & 0 & 0 & 0 & 0 & 0 & 0 & 0 & 0 & 0 & 0 & 0 & 0 & 0 & 0 & 0 & 0 & 0 & 0 & 0 & 0 & 0 & 0 & 0 & 0 & 0 & 0 & 0 & 0 & 0 & 0 & 0 & 0 & 0 & 0 & 0 & 0 & 0 & 0 & 0 & 0 & 0 & 0 & 0 & 0 & 0 & 0 & 0 & 0 & 0 & 0 & 0 & 1 & 0 \\
                0 & 0 & 0 & 0 & 0 & 0 & 0 & 0 & 0 & 0 & 0 & 0 & 0 & 0 & 0 & 0 & 0 & 0 & 0 & 0 & 0 & 0 & 0 & 0 & 0 & 0 & 0 & 0 & 0 & 0 & 0 & 0 & 0 & 0 & 0 & 0 & 0 & 0 & 0 & 0 & 0 & 0 & 0 & 0 & 0 & 0 & 0 & 0 & 0 & 0 & 0 & 0 & 0 & 0 & 0 & 0 & 0 & 0 & 0 & 0 & 0 & 1 & 0 & 0 \\
                0 & 0 & 0 & 0 & 0 & 0 & 0 & 0 & 0 & 0 & 0 & 0 & 0 & 0 & 0 & 0 & 0 & 0 & 0 & 0 & 0 & 0 & 0 & 0 & 0 & 0 & 0 & 0 & 0 & 0 & 0 & 0 & 0 & 0 & 0 & 0 & 0 & 0 & 0 & 0 & 0 & 0 & 0 & 0 & 0 & 0 & 0 & 0 & 0 & 0 & 0 & 0 & 0 & 0 & 0 & 0 & 0 & 0 & 0 & 0 & 1 & 0 & 0 & 0 \\
                0 & 0 & 0 & 0 & 0 & 0 & 0 & 0 & 0 & 0 & 0 & 0 & 0 & 0 & 0 & 0 & 0 & 0 & 0 & 0 & 0 & 0 & 0 & 0 & 0 & 0 & 0 & 0 & 0 & 0 & 0 & 0 & 0 & 0 & 0 & 0 & 0 & 0 & 0 & 0 & 0 & 0 & 0 & 0 & 0 & 0 & 0 & 0 & 0 & 0 & 0 & 0 & 0 & 0 & 0 & 0 & 0 & 0 & 0 & 1 & 0 & 0 & 0 & 0 \\
                0 & 0 & 0 & 0 & 0 & 0 & 0 & 0 & 0 & 0 & 0 & 0 & 0 & 0 & 0 & 0 & 0 & 0 & 0 & 0 & 0 & 0 & 0 & 0 & 0 & 0 & 0 & 0 & 0 & 0 & 0 & 0 & 0 & 0 & 0 & 0 & 0 & 0 & 0 & 0 & 0 & 0 & 0 & 0 & 0 & 0 & 0 & 0 & 0 & 0 & 0 & 0 & 0 & 0 & 0 & 0 & 0 & 0 & 1 & 0 & 0 & 0 & 0 & 0 \\
                0 & 0 & 0 & 0 & 0 & 0 & 0 & 0 & 0 & 0 & 0 & 0 & 0 & 0 & 0 & 0 & 0 & 0 & 0 & 0 & 0 & 0 & 0 & 0 & 0 & 0 & 0 & 0 & 0 & 0 & 0 & 0 & 0 & 0 & 0 & 0 & 0 & 0 & 0 & 0 & 0 & 0 & 0 & 0 & 0 & 0 & 0 & 0 & 0 & 0 & 0 & 0 & 0 & 0 & 0 & 0 & 0 & 1 & 0 & 0 & 0 & 0 & 0 & 0 \\
                0 & 0 & 0 & 0 & 0 & 0 & 0 & 0 & 0 & 0 & 0 & 0 & 0 & 0 & 0 & 0 & 0 & 0 & 0 & 0 & 0 & 0 & 0 & 0 & 0 & 0 & 0 & 0 & 0 & 0 & 0 & 0 & 0 & 0 & 0 & 0 & 0 & 0 & 0 & 0 & 0 & 0 & 0 & 0 & 0 & 0 & 0 & 0 & 0 & 0 & 0 & 0 & 0 & 0 & 0 & 0 & 1 & 0 & 0 & 0 & 0 & 0 & 0 & 0
           \end{pmatrix}$
       }
   }
\end{center}

% SSP Qid 6 Conv
\clearpage
The unitary matrix for \QidConvSSP.
This gate is used to convert \ndimensional{6} qudits to qubits when analyzing \gls{ssp}.
\begin{center}
     \makebox[\textwidth]{%
%     \resizebox{0.99\paperwidth}{!}{%
          $\begin{pmatrix}
        1 & 0 & 0 & 0 & 0 & 0 & 0 & 0 & 0 & 0 & 0 & 0 & 0 & 0 & 0 & 0 & 0 & 0 & 0 & 0 & 0 & 0 & 0 & 0 & 0 & 0 & 0 & 0 & 0 & 0 & 0 & 0 & 0 & 0 & 0 & 0 \\
        0 & 1 & 0 & 0 & 0 & 0 & 0 & 0 & 0 & 0 & 0 & 0 & 0 & 0 & 0 & 0 & 0 & 0 & 0 & 0 & 0 & 0 & 0 & 0 & 0 & 0 & 0 & 0 & 0 & 0 & 0 & 0 & 0 & 0 & 0 & 0 \\
        0 & 0 & 1 & 0 & 0 & 0 & 0 & 0 & 0 & 0 & 0 & 0 & 0 & 0 & 0 & 0 & 0 & 0 & 0 & 0 & 0 & 0 & 0 & 0 & 0 & 0 & 0 & 0 & 0 & 0 & 0 & 0 & 0 & 0 & 0 & 0 \\
        0 & 0 & 0 & 1 & 0 & 0 & 0 & 0 & 0 & 0 & 0 & 0 & 0 & 0 & 0 & 0 & 0 & 0 & 0 & 0 & 0 & 0 & 0 & 0 & 0 & 0 & 0 & 0 & 0 & 0 & 0 & 0 & 0 & 0 & 0 & 0 \\
        0 & 0 & 0 & 0 & 1 & 0 & 0 & 0 & 0 & 0 & 0 & 0 & 0 & 0 & 0 & 0 & 0 & 0 & 0 & 0 & 0 & 0 & 0 & 0 & 0 & 0 & 0 & 0 & 0 & 0 & 0 & 0 & 0 & 0 & 0 & 0 \\
        0 & 0 & 0 & 0 & 0 & 1 & 0 & 0 & 0 & 0 & 0 & 0 & 0 & 0 & 0 & 0 & 0 & 0 & 0 & 0 & 0 & 0 & 0 & 0 & 0 & 0 & 0 & 0 & 0 & 0 & 0 & 0 & 0 & 0 & 0 & 0 \\
        0 & 0 & 0 & 0 & 0 & 0 & 0 & 0 & 1 & 0 & 0 & 0 & 0 & 0 & 0 & 0 & 0 & 0 & 0 & 0 & 0 & 0 & 0 & 0 & 0 & 0 & 0 & 0 & 0 & 0 & 0 & 0 & 0 & 0 & 0 & 0 \\
        0 & 0 & 0 & 0 & 0 & 0 & 0 & 0 & 0 & 1 & 0 & 0 & 0 & 0 & 0 & 0 & 0 & 0 & 0 & 0 & 0 & 0 & 0 & 0 & 0 & 0 & 0 & 0 & 0 & 0 & 0 & 0 & 0 & 0 & 0 & 0 \\
        0 & 0 & 0 & 0 & 0 & 0 & 1 & 0 & 0 & 0 & 0 & 0 & 0 & 0 & 0 & 0 & 0 & 0 & 0 & 0 & 0 & 0 & 0 & 0 & 0 & 0 & 0 & 0 & 0 & 0 & 0 & 0 & 0 & 0 & 0 & 0 \\
        0 & 0 & 0 & 0 & 0 & 0 & 0 & 1 & 0 & 0 & 0 & 0 & 0 & 0 & 0 & 0 & 0 & 0 & 0 & 0 & 0 & 0 & 0 & 0 & 0 & 0 & 0 & 0 & 0 & 0 & 0 & 0 & 0 & 0 & 0 & 0 \\
        0 & 0 & 0 & 0 & 0 & 0 & 0 & 0 & 0 & 0 & 0 & 1 & 0 & 0 & 0 & 0 & 0 & 0 & 0 & 0 & 0 & 0 & 0 & 0 & 0 & 0 & 0 & 0 & 0 & 0 & 0 & 0 & 0 & 0 & 0 & 0 \\
        0 & 0 & 0 & 0 & 0 & 0 & 0 & 0 & 0 & 0 & 1 & 0 & 0 & 0 & 0 & 0 & 0 & 0 & 0 & 0 & 0 & 0 & 0 & 0 & 0 & 0 & 0 & 0 & 0 & 0 & 0 & 0 & 0 & 0 & 0 & 0 \\
        0 & 0 & 0 & 0 & 0 & 0 & 0 & 0 & 0 & 0 & 0 & 0 & 0 & 0 & 0 & 0 & 1 & 0 & 0 & 0 & 0 & 0 & 0 & 0 & 0 & 0 & 0 & 0 & 0 & 0 & 0 & 0 & 0 & 0 & 0 & 0 \\
        0 & 0 & 0 & 0 & 0 & 0 & 0 & 0 & 0 & 0 & 0 & 0 & 0 & 0 & 0 & 0 & 0 & 1 & 0 & 0 & 0 & 0 & 0 & 0 & 0 & 0 & 0 & 0 & 0 & 0 & 0 & 0 & 0 & 0 & 0 & 0 \\
        0 & 0 & 0 & 0 & 0 & 0 & 0 & 0 & 0 & 0 & 0 & 0 & 0 & 0 & 0 & 1 & 0 & 0 & 0 & 0 & 0 & 0 & 0 & 0 & 0 & 0 & 0 & 0 & 0 & 0 & 0 & 0 & 0 & 0 & 0 & 0 \\
        0 & 0 & 0 & 0 & 0 & 0 & 0 & 0 & 0 & 0 & 0 & 0 & 0 & 0 & 1 & 0 & 0 & 0 & 0 & 0 & 0 & 0 & 0 & 0 & 0 & 0 & 0 & 0 & 0 & 0 & 0 & 0 & 0 & 0 & 0 & 0 \\
        0 & 0 & 0 & 0 & 0 & 0 & 0 & 0 & 0 & 0 & 0 & 0 & 1 & 0 & 0 & 0 & 0 & 0 & 0 & 0 & 0 & 0 & 0 & 0 & 0 & 0 & 0 & 0 & 0 & 0 & 0 & 0 & 0 & 0 & 0 & 0 \\
        0 & 0 & 0 & 0 & 0 & 0 & 0 & 0 & 0 & 0 & 0 & 0 & 0 & 1 & 0 & 0 & 0 & 0 & 0 & 0 & 0 & 0 & 0 & 0 & 0 & 0 & 0 & 0 & 0 & 0 & 0 & 0 & 0 & 0 & 0 & 0 \\
        0 & 0 & 0 & 0 & 0 & 0 & 0 & 0 & 0 & 0 & 0 & 0 & 0 & 0 & 0 & 0 & 0 & 0 & 0 & 1 & 0 & 0 & 0 & 0 & 0 & 0 & 0 & 0 & 0 & 0 & 0 & 0 & 0 & 0 & 0 & 0 \\
        0 & 0 & 0 & 0 & 0 & 0 & 0 & 0 & 0 & 0 & 0 & 0 & 0 & 0 & 0 & 0 & 0 & 0 & 1 & 0 & 0 & 0 & 0 & 0 & 0 & 0 & 0 & 0 & 0 & 0 & 0 & 0 & 0 & 0 & 0 & 0 \\
        0 & 0 & 0 & 0 & 0 & 0 & 0 & 0 & 0 & 0 & 0 & 0 & 0 & 0 & 0 & 0 & 0 & 0 & 0 & 0 & 0 & 1 & 0 & 0 & 0 & 0 & 0 & 0 & 0 & 0 & 0 & 0 & 0 & 0 & 0 & 0 \\
        0 & 0 & 0 & 0 & 0 & 0 & 0 & 0 & 0 & 0 & 0 & 0 & 0 & 0 & 0 & 0 & 0 & 0 & 0 & 0 & 1 & 0 & 0 & 0 & 0 & 0 & 0 & 0 & 0 & 0 & 0 & 0 & 0 & 0 & 0 & 0 \\
        0 & 0 & 0 & 0 & 0 & 0 & 0 & 0 & 0 & 0 & 0 & 0 & 0 & 0 & 0 & 0 & 0 & 0 & 0 & 0 & 0 & 0 & 0 & 1 & 0 & 0 & 0 & 0 & 0 & 0 & 0 & 0 & 0 & 0 & 0 & 0 \\
        0 & 0 & 0 & 0 & 0 & 0 & 0 & 0 & 0 & 0 & 0 & 0 & 0 & 0 & 0 & 0 & 0 & 0 & 0 & 0 & 0 & 0 & 1 & 0 & 0 & 0 & 0 & 0 & 0 & 0 & 0 & 0 & 0 & 0 & 0 & 0 \\
        0 & 0 & 0 & 0 & 0 & 0 & 0 & 0 & 0 & 0 & 0 & 0 & 0 & 0 & 0 & 0 & 0 & 0 & 0 & 0 & 0 & 0 & 0 & 0 & 0 & 0 & 0 & 1 & 0 & 0 & 0 & 0 & 0 & 0 & 0 & 0 \\
        0 & 0 & 0 & 0 & 0 & 0 & 0 & 0 & 0 & 0 & 0 & 0 & 0 & 0 & 0 & 0 & 0 & 0 & 0 & 0 & 0 & 0 & 0 & 0 & 0 & 0 & 1 & 0 & 0 & 0 & 0 & 0 & 0 & 0 & 0 & 0 \\
        0 & 0 & 0 & 0 & 0 & 0 & 0 & 0 & 0 & 0 & 0 & 0 & 0 & 0 & 0 & 0 & 0 & 0 & 0 & 0 & 0 & 0 & 0 & 0 & 0 & 1 & 0 & 0 & 0 & 0 & 0 & 0 & 0 & 0 & 0 & 0 \\
        0 & 0 & 0 & 0 & 0 & 0 & 0 & 0 & 0 & 0 & 0 & 0 & 0 & 0 & 0 & 0 & 0 & 0 & 0 & 0 & 0 & 0 & 0 & 0 & 1 & 0 & 0 & 0 & 0 & 0 & 0 & 0 & 0 & 0 & 0 & 0 \\
        0 & 0 & 0 & 0 & 0 & 0 & 0 & 0 & 0 & 0 & 0 & 0 & 0 & 0 & 0 & 0 & 0 & 0 & 0 & 0 & 0 & 0 & 0 & 0 & 0 & 0 & 0 & 0 & 0 & 1 & 0 & 0 & 0 & 0 & 0 & 0 \\
        0 & 0 & 0 & 0 & 0 & 0 & 0 & 0 & 0 & 0 & 0 & 0 & 0 & 0 & 0 & 0 & 0 & 0 & 0 & 0 & 0 & 0 & 0 & 0 & 0 & 0 & 0 & 0 & 1 & 0 & 0 & 0 & 0 & 0 & 0 & 0 \\
        0 & 0 & 0 & 0 & 0 & 0 & 0 & 0 & 0 & 0 & 0 & 0 & 0 & 0 & 0 & 0 & 0 & 0 & 0 & 0 & 0 & 0 & 0 & 0 & 0 & 0 & 0 & 0 & 0 & 0 & 0 & 0 & 0 & 0 & 0 & 1 \\
        0 & 0 & 0 & 0 & 0 & 0 & 0 & 0 & 0 & 0 & 0 & 0 & 0 & 0 & 0 & 0 & 0 & 0 & 0 & 0 & 0 & 0 & 0 & 0 & 0 & 0 & 0 & 0 & 0 & 0 & 0 & 0 & 0 & 0 & 1 & 0 \\
        0 & 0 & 0 & 0 & 0 & 0 & 0 & 0 & 0 & 0 & 0 & 0 & 0 & 0 & 0 & 0 & 0 & 0 & 0 & 0 & 0 & 0 & 0 & 0 & 0 & 0 & 0 & 0 & 0 & 0 & 0 & 0 & 0 & 1 & 0 & 0 \\
        0 & 0 & 0 & 0 & 0 & 0 & 0 & 0 & 0 & 0 & 0 & 0 & 0 & 0 & 0 & 0 & 0 & 0 & 0 & 0 & 0 & 0 & 0 & 0 & 0 & 0 & 0 & 0 & 0 & 0 & 0 & 0 & 1 & 0 & 0 & 0 \\
        0 & 0 & 0 & 0 & 0 & 0 & 0 & 0 & 0 & 0 & 0 & 0 & 0 & 0 & 0 & 0 & 0 & 0 & 0 & 0 & 0 & 0 & 0 & 0 & 0 & 0 & 0 & 0 & 0 & 0 & 0 & 1 & 0 & 0 & 0 & 0 \\
        0 & 0 & 0 & 0 & 0 & 0 & 0 & 0 & 0 & 0 & 0 & 0 & 0 & 0 & 0 & 0 & 0 & 0 & 0 & 0 & 0 & 0 & 0 & 0 & 0 & 0 & 0 & 0 & 0 & 0 & 1 & 0 & 0 & 0 & 0 & 0
   \end{pmatrix}$
%          }
     }
\end{center}